\documentclass[aps,preprint,showpacs,groupedaddress,nofootinbib]{revtex4}

\usepackage{graphicx}
\usepackage{axodraw}
\usepackage{color}

\def\red#1{{\color{red}#1}}
\def\green#1{{\color{green}#1}}
\def\blue#1{{\color{blue}#1}}
\newcommand{\met}{\rlap{\,/}E_T}

\begin{document}

\preprint{MCTP-05-93}
\preprint{UFIFT-HEP-05-14}

\vspace*{2cm}

\title{Discrimination of Supersymmetry and Universal Extra 
Dimensions at Hadron Colliders}

\author{AseshKrishna Datta}
\affiliation{MCTP, University of Michigan, Ann Arbor, MI 48109-1120, USA}
\author{Kyoungchul Kong}
\affiliation{Institute for Fundamental Theory, Physics Department, University of Florida, Gainesville, FL 32611, USA}
\author{Konstantin T.~Matchev}
\affiliation{Institute for Fundamental Theory, Physics Department, University of Florida, Gainesville, FL 32611, USA}

\date{September 22, 2005}

\begin{abstract}
We contrast the experimental signatures of low energy supersymmetry 
and the model of Universal Extra Dimensions and discuss various
methods for their discrimination at hadron colliders.
We study the discovery reach of the Tevatron and the LHC for level 2 
Kaluza-Klein modes, which would indicate the presence of extra dimensions.
We find that with 100 ${\rm fb}^{-1}$ of data the LHC will be able to 
discover the $\gamma_2$ and $Z_2$ KK modes as separate resonances
if their masses are below 2 TeV.
We also investigate the possibility to differentiate the spins 
of the superpartners and KK modes by means of the asymmetry method of Barr.
\end{abstract}


\maketitle

\section{\label{sec:intro}Introduction}

With the highly anticipated run of the Large Hadron Collider (LHC) at CERN
we will begin to explore the Terascale in earnest. There are very sound 
reasons to expect momentous discoveries at the LHC. Among the greatest 
mysteries in particle physics today is the origin of electroweak 
symmetry breaking, which, according to the Standard Model, is accomplished 
through the Higgs mechanism. The Higgs particle is the primary target
of the LHC experiments and, barring some unexpected behavior, the Higgs
boson will be firmly discovered after only a few years of running of the LHC. 
With some luck, a Higgs signal might start appearing already in the Tevatron 
Run II. 

The discovery of a Higgs boson, however, will open a host of new questions.
As the first fundamental scalar to be seen, it will bring about a 
worrisome fine tuning problem: why is the Higgs particle so light, 
compared to, say, the Planck scale? Various solutions to this hierarchy 
problem have been proposed, and the most aesthetically pleasing one at this point 
appears to be low energy supersymmetry (SUSY). In SUSY, the problematic 
quadratic divergences in the radiative corrections to the 
Higgs mass are absent, being canceled by loops with superpartners.
The cancellations are enforced by the symmetry, and the Higgs mass
is therefore naturally related to the mass scale of the superpartners.

While the solution of the hierarchy problem is perhaps
the most celebrated virtue of SUSY, supersymmetric models have 
other side benefits. For one, if the superpartners are 
indeed within the TeV range, they would modify the running of 
the gauge couplings at higher scales, and gauge coupling unification
takes place with astonishing precision. Secondly, a large 
class of SUSY models, which have a conserved discrete symmetry ($R$-parity),
contain an excellent dark matter candidate:
the lightest neutralino $\tilde\chi^0_1$. 
One should keep in mind that the dark matter problem is by far 
the most compelling {\em experimental} evidence for particles 
and interactions outside the Standard Model (SM), and provides a completely 
independent motivation for entertaining supersymmetry at the TeV scale. 
Finally, $R$-parity implies that superpartners interact
only pairwise with SM particles, which guarantees that the 
supersymmetric contributions to low energy precision data 
only appear at the loop level and are small. In summary,
supersymmetric extensions of the SM are the primary candidates
for new physics at the TeV scale. Not surprisingly, therefore,
the signatures of supersymmetry at the Tevatron and LHC have been
extensively discussed in the literature. In typical scenarios
with superpartners in the range of a few TeV or less, already
within the first few years of running the LHC will discover a signal 
of new physics in several channels. Once such a signal of physics beyond 
the Standard Model is seen, it will immediately bring up the question:
is it supersymmetry or not?

The answer to this question can be approached in two different ways.
On the theoretical side, one may ask whether there are well 
motivated alternatives to low energy supersymmetry, which would
give similar signatures at hadron colliders, in other words, if the new 
physics is not supersymmetry, what else can it be? Until recently,
there were no known examples of other types of new physics, which 
could ``fake'' supersymmetry sufficiently well. The signatures of 
supersymmetry and its competitors (technicolor, new gauge bosons,
large extra dimensions, etc.) were sufficiently distinctive, and 
there was little room for confusion. However, it was recently 
realized that the framework of Universal Extra Dimensions (UED), originally 
proposed in~\cite{Appelquist:2000nn}, can very effectively masquerade 
as low energy supersymmetry at a hadron collider such as the LHC 
or the Tevatron~\cite{Cheng:2002ab}. It therefore became of sufficient
interest to try and prove supersymmetry at the LHC from first principles,
without resorting to model-dependent assumptions and without theoretical 
bias. The experimental program for proving supersymmetry at a {\em lepton}
collider has been outlined a long time ago~\cite{Feng:1995zd} and can be 
readily followed to make the discrimination between SUSY and 
UED~\cite{Battaglia:2005zf,Bhattacharyya:2005vm,Bhattacherjee:2005qe,Riemann:2005es}. 
However, as we shall see below, 
the case of hadron colliders is much more challenging.

The more direct approach to confirming supersymmetry at the LHC 
would be to first ask: what are the defining features of supersymmetry, 
and can we prove them from the data? By now there is a wide variety of 
supersymmetric models, with very diverse phenomenology. Nevertheless,
they all share the following common features which define a supersymmetric 
framework:
\begin{enumerate}
\item For each particle of the Standard Model, supersymmetry predicts
a new particle (superpartner).
\item The spins of the superpartners differ by $1/2$ unit.
\item The couplings of the particles and their superpartners are
equal, being related by supersymmetry.
\end{enumerate}
If supersymmetry were exact, one would have another common feature:
the prediction that the masses of the particles and their superpartners 
must be equal as well. However, once SUSY 
is broken (as it must be), the superpartner masses are lifted and 
one obtains spectra classified by the mechanism of SUSY breaking:
supergravity-mediated, gauge-mediated, gaugino-mediated, 
anomaly-mediated etc.~(for a recent review, see~\cite{Chung:2003fi}). 
As a result, in realistic models of low-energy 
supersymmetry the pattern of sparticle masses is very 
model-dependent. The measurements of the superpartner masses
are therefore probing SUSY {\em breaking} phenomena rather than
a fundamental property of supersymmetry itself.

In the following let us only concentrate on SUSY models which possess
a weakly interacting massive particle (WIMP) as a 
dark matter candidate, as guaranteed by $R$-parity
conservation. This will be in fact the most difficult case for
establishing supersymmetry at a hadron collider. Due to $R$-parity, 
the superpartners are pair-produced, and each one decays to the lightest
supersymmetric particle (LSP), in this case $\tilde\chi^0_1$.
Since the two $\tilde\chi^0_1$s leave the detector without any 
interaction, the generic SUSY signal at the LHC is missing energy.
With these assumptions, we can add another identifying feature 
to the list above, although this is not required by supersymmetry
itself:
\begin{enumerate}
\setcounter{enumi}{3}
\item The generic collider signature of supersymmetric models with 
WIMP LSPs is missing energy.
\end{enumerate}
This last property makes exact reconstruction of the event kinematics 
practically impossible. At a hadron collider, the center of mass energy 
is not known on an event per event basis. In addition, the momenta 
of {\em both} $\tilde\chi^0_1$ particles are unknown, and what is 
measured is only the transverse component of the sum of their momenta, 
provided there are no other sources of missing energy in the event (such as 
neutrinos, $b$-jets, $\tau$-jets, etc.). As we shall see below, this  
incomplete information is the main stumbling block in proving the basic
properties of supersymmetry at the LHC. 

The purpose of this paper is to investigate the prospects for
establishing supersymmetry at the LHC by discriminating it from its 
look-alike scenario of Universal Extra Dimensions.
In Section~\ref{sec:ued} we review the basic phenomenology of the UED
model, contrasting it with a generic supersymmetric model as described above. 
We identify two basic discriminators between the two scenarios and 
proceed to consider each one in turn in the following two sections. 
One of the characteristic features of extra dimensional models is the presence
of a whole tower of Kaluza-Klein (KK) partners, labelled by their KK level $n$.
In contrast, $N=1$ supersymmetry predicts a single superpartner for each SM particle.
One might therefore hope to discover the higher KK modes of UED
and thus prove the existence of extra dimensions. 
In Section~\ref{sec:level2} we study the discovery reach
for level 2 KK gauge boson particles and the resolving power of
the LHC to see them as separate resonances.
The other fundamental difference between SUSY and UED is the spin
of the superpartners (KK partners). In Section~\ref{sec:spin}
we study how well the two models can be distinguished based on spin 
correlations in the cascade decays of the new particles. In particular, we
use the asymmetry variable recently advertised by Barr~\cite{Barr:2004ze}
for this purpose\footnote{While this work was in 
preparation~\cite{KongLCWS,KMAPS,KongPheno,Battaglia:2005ma}, 
a similar study of the asymmetry was published in~\cite{Smillie:2005ar},
with very similar conclusions, see Section~\ref{sec:spin}.}.
We present our conclusions in Section~\ref{sec:conclusions}.

\section{\label{sec:ued}Phenomenology of Universal Extra Dimensions}

\subsection{The Minimal UED Model}
\label{sec:MUED}

Models of UED place all Standard Model particles in the bulk of 
one or more compactified extra dimensions. In the simplest and most popular
version, there is a single extra dimension of size $R$,
compactified on an $S_1/Z_2$ orbifold~\cite{Appelquist:2000nn}.
More complicated 6-dimensional models have also been 
built~\cite{Dobrescu:2004zi,Burdman:2005sr}.
The UED framework has been a fruitful playground for
addressing different
puzzles of the Standard Model, such as electroweak symmetry 
breaking and vacuum stability~\cite{Arkani-Hamed:2000hv,Bucci:2003fk,Bucci:2004xw},
neutrino masses~\cite{Appelquist:2002ft,Mohapatra:2002ug}, 
proton stability~\cite{Appelquist:2001mj} or the number of 
generations~\cite{Dobrescu:2001ae}. A peculiar feature of UED 
is the conservation of
Kaluza-Klein number at tree level, which is a simple consequence
of momentum conservation along the extra dimension.
However, bulk and brane radiative effects 
\cite{Georgi:2000ks,vonGersdorff:2002as,Cheng:2002iz}
break KK number down to a discrete conserved quantity,
the so called KK parity, $(-1)^n$, where $n$ is the KK level.
KK parity ensures that the lightest KK partners 
(those at level one) are always pair-produced
in collider experiments, just like in the $R$-parity conserving 
supersymmetry models discussed in Section~\ref{sec:intro}.
KK parity conservation also implies
that the contributions to various low-energy
observables \cite{Agashe:2001ra,Agashe:2001xt,Appelquist:2001jz,%
Petriello:2002uu,Appelquist:2002wb,Chakraverty:2002qk,Buras:2002ej,%
Oliver:2002up,Buras:2003mk,Iltan:2003tn,Khalil:2004qk}
only arise at loop level and are small.
As a result, the limits on the scale $R^{-1}$ of the extra dimension
from precision electroweak data are rather weak, constraining  $R^{-1}$ to 
be larger than approximately 250~GeV \cite{Appelquist:2002wb}. 
An attractive feature of UED models with KK parity 
is the presence of a stable massive particle which can be
a cold dark matter candidate \cite{Dienes:1998vg,Cheng:2002iz}.
The lightest KK partner (LKP) at level one 
is also the lightest particle with negative KK parity and 
is stable on cosmological scales.
The identity of the LKP is a delicate issue, however, as it
depends on the interplay between the one-loop radiative corrections
to the KK mass spectrum and the brane terms generated by unknown physics 
at high scales~\cite{Cheng:2002iz}.
In the minimal UED model defined below, 
the LKP turns out to be the KK partner $B_1$ of the
hypercharge gauge boson \cite{Cheng:2002iz} and
its relic density is typically in the right ballpark:
in order to explain all of the dark matter, the $B_1$ mass
should be in the range 500-600 
GeV~\cite{Servant:2002aq,Majumdar:2003dj,Kakizaki:2005en,Kakizaki:2005uy,Burnell:2005hm,Kong:2005hn}. 
Kaluza-Klein dark matter offers 
excellent prospects for direct~\cite{Cheng:2002ej,Servant:2002hb,Majumdar:2002mw}
or indirect detection~\cite{Cheng:2002ej,Hooper:2002gs,Bertone:2002ms,Hooper:2004xn,%
Bergstrom:2004cy,Baltz:2004ie,Bergstrom:2004nr,Bringmann:2005pp,Barrau:2005au,Birkedal:2005ep}.
Once the radiative corrections to the Kaluza-Klein masses are 
properly taken into account,
the collider phenomenology of the Minimal UED model
exhibits striking similarities to supersymmetry
\cite{Cheng:2002ab,Cheng:2002rn} and represents 
an interesting and well motivated counterexample 
which can ``fake'' supersymmetry signals at the LHC.

For the purposes of our study we have chosen to work with the minimal 
UED model considered in~\cite{Cheng:2002ab}.
In UED the bulk interactions of the KK modes are fixed by
the SM Lagrangian and contain no unknown parameters other
than the mass, $m_h$, of the SM Higgs boson. In contrast, the boundary
interactions, which are localized on the orbifold fixed points, are in 
principle arbitrary, and their coefficients represent new free parameters 
in the theory. Since the boundary terms are renormalized
by bulk interactions, they are scale dependent \cite{Georgi:2000ks}
and cannot be completely ignored since they will be generated by 
renormalization effects. 
Therefore, one needs an ansatz for their values at a particular scale.
Like any higher dimensional Kaluza-Klein theory, the UED model 
should be treated only as an effective theory valid up to some 
high scale $\Lambda$, at which it matches to some more fundamental theory. 
The Minimal UED model is then defined so that the coefficients of all
boundary interactions vanish at this matching scale $\Lambda$, but are
subsequently generated through RGE evolution to lower scales. 
The Minimal UED model therefore has only two input parameters: 
the size of the extra dimension, $R$, and the cutoff scale, $\Lambda$. 
The number of KK levels present in the effective theory is 
simply $\Lambda R$ and may vary between a few and $\sim 40$, 
where the upper limit comes from the breakdown of perturbativity 
already below the scale $\Lambda$. Unless specified otherwise,
for our numerical results below, we shall always choose the 
value of $\Lambda$ so that $\Lambda R=20$. Changing the value 
of $\Lambda$ will have very little impact on our results since the 
$\Lambda$ dependence of the KK mass spectrum is only logarithmic.

\begin{figure}[t]
\includegraphics[width=10cm]{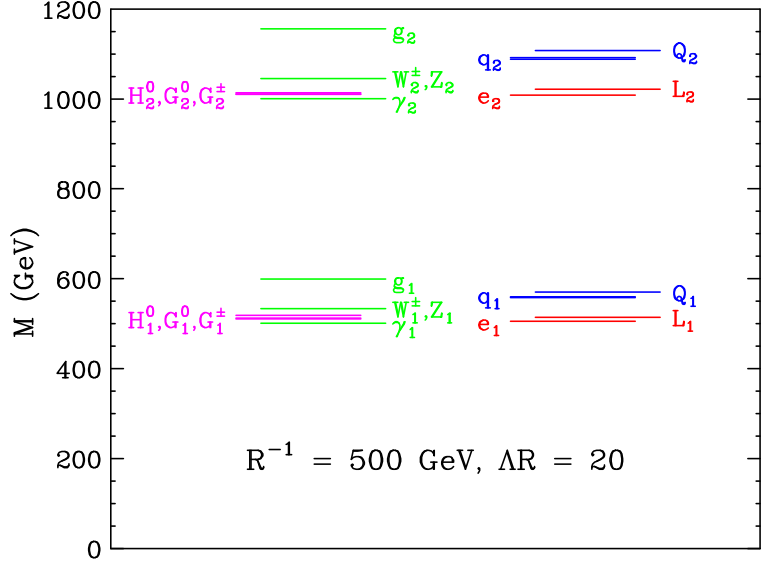}
\caption{One-loop corrected mass spectrum of the $n=1$ and $n=2$ KK levels
in Minimal UED, for $R^{-1}=500$ GeV, $\Lambda R=20$ and $m_h=120$ GeV.
We show the KK modes of gauge bosons, Higgs and Goldstone bosons
and first generation fermions.
\label{fig:spectrum}}
\end{figure}

In Fig.~\ref{fig:spectrum} we show the mass spectrum of the $n=1$ and 
$n=2$ KK levels in Minimal UED, for $R^{-1}=500$ GeV, 
$\Lambda R=20$ and SM Higgs boson mass $m_h=120$ GeV. 
We include the full one-loop corrections 
from \cite{Cheng:2002iz}. We have used RGE improved 
couplings to compute the radiative corrections to the KK masses. 
It is well known that in UED the KK modes modify the running 
of the coupling constants at higher scales. 
We extrapolate the gauge coupling constants to
the scale of the $n=1$ and $n=2$ KK modes, using 
the appropriate $\beta$ functions dictated by the particle 
spectrum~\cite{Dienes:1998vg,Perez-Lorenzana:1999qb,Cheng:1999fu}.
As a result the spectrum shown in Fig.~\ref{fig:spectrum}
differs slightly from the one in \cite{Cheng:2002iz}.
Most notably, the colored KK particles are somewhat lighter, due 
to a reduced value of the strong coupling constant, and overall 
the KK spectrum at each level is more degenerate.

\subsection{Comparison of UED and Supersymmetry}

We are now in a position to compare in general terms the
phenomenology of UED and supersymmetry at colliders. 
In Section~\ref{sec:intro} we outlined four identifying features of 
SUSY models with WIMP LSPs. In complete analogy, the discussion of 
Section~\ref{sec:MUED} leads to the following generic features of UED:
\begin{enumerate}
\item For each particle of the Standard Model, UED models predict
an infinite\footnote{Strictly speaking, the number of KK modes 
is $\Lambda R$, see Section~\ref{sec:MUED}.} tower of 
new particles (Kaluza-Klein partners).
\item The spins of the SM particles and their KK partners are the same.
\item The couplings of the SM particles and their KK partners are equal.
\item The generic collider signature of UED models with 
WIMP LKPs is missing energy.
\end{enumerate}
Notice that the defining features 3 and 4 are common to both supersymmetry 
and UED and cannot be used to distinguish the two cases.
We see that while $R$-parity conserving SUSY implies 
a missing energy signal, the reverse is not true: a missing energy
signal would appear in any model with a dark matter candidate, 
and even in models which have nothing to do with the dark 
matter issue, but simply contain new neutral quasi-stable particles,
e.g.~gravitons~\cite{Arkani-Hamed:1998rs,Giudice:1998ck,Mirabelli:1998rt}. 
Similarly, the equality of the couplings 
(feature No.~3) is a celebrated test of SUSY, but from the above comparison 
we see that it is only a necessary, but not a sufficient condition in 
proving supersymmetry. In addition, the measurement of superpartner 
couplings in order to test the SUSY relations is a very challenging 
task at a hadron collider. For one, the observed production rate 
in any given channel is only sensitive to the product
of the cross-section times the branching fractions, and so any
attempt to measure the couplings from a cross-section 
would have to make certain assumptions about the branching fractions.
An additional complication arises from the fact that at hadron colliders all
kinematically available states can be produced simultaneously, and 
the production of a particular species in an exclusive channel 
is rather difficult to isolate. The couplings could also in principle 
be measured from the branching fractions, but that also requires 
a measurement of the total width, which is impossible in our case, 
since the Breit-Wigner resonance cannot be reconstructed, 
due to the unknown momentum of the missing LSP (LKP).

We are therefore forced to concentrate on the first two identifying
features as the only promising discriminating criteria. 
Let us begin with feature 1: the number of new particles.
The KK particles at $n=1$ are analogous to superpartners 
in supersymmetry. However, the particles at the higher 
KK levels have no analogues in $N=1$ supersymmetric models.
Discovering the $n\ge2$ levels of the KK tower would therefore 
indicate the presence of extra dimensions rather than SUSY.
In this study we shall concentrate on the $n=2$ level and
in Section~\ref{sec:level2} we investigate the discovery 
opportunities at the LHC and the Tevatron (for linear collider 
studies of $n=2$ KK gauge bosons, 
see \cite{Battaglia:2005zf,Bhattacherjee:2005qe,Riemann:2005es,Battaglia:2005ma}).
Notice that the masses of the KK modes are given roughly by
$m_n\sim n/R$, where $n$ is the KK level number, so that the particles 
at levels 3 and higher are rather heavy and their production
is severely suppressed. 

The second identifying feature -- the spins of the new particles -- 
also provides a tool for discrimination between SUSY and UED:
the KK partners have identical spin quantum numbers as their SM 
counterparts, while the spins of the superpartners differ by $1/2$ unit. 
However, spin determinations are known to be difficult at the LHC
(or at hadron colliders in general), where the parton-level 
center of mass energy $E_{CM}$ in each event is unknown. 
In addition, the momenta of the two dark matter candidates 
in the event are also unknown. This prevents the reconstruction 
of any rest frame angular decay distributions, or the directions 
of the two particles at the top of the decay chains. 
The variable $E_{CM}$ also rules out the possibility of a threshold scan, 
which is one of the main tools for determining particle spins at 
lepton colliders. We are therefore forced to look for new methods for
spin determinations, or at least for finding spin 
correlations\footnote{Notice that in simple processes with
two-body decays like slepton production $e^+e^-\to \tilde\mu^+\tilde\mu^-
\to \mu^+\mu^-\tilde\chi^0_1\tilde\chi^0_1$
the flat energy distribution of the observable final state particles 
(muons in this case) is often regarded as a smoking gun for the
scalar nature of the intermediate particles (the smuons).
Indeed, the smuons are spin zero particles and decay isotropically 
in their rest frame, which results in a flat distribution in the lab frame.
However, the flat distribution is a necessary but not sufficient 
condition for a scalar particle, and UED provides a counterexample
with the analogous process of KK muon production~\cite{Battaglia:2005zf},
where a flat distribution also appears, but as a result of equal contributions 
from left-handed and right-handed KK fermions.}. Recently it has been
suggested that a charge asymmetry in the lepton-jet invariant mass
distributions from a particular cascade, can be used to discriminate
SUSY from the case of pure phase space decays~\cite{Barr:2004ze}.
The possibility of discriminating SUSY and UED by this method 
will be the subject of Section~\ref{sec:spin} (see 
also~\cite{KongLCWS,KMAPS,KongPheno,Battaglia:2005ma}
and ~\cite{Smillie:2005ar}).

For the purposes of our study we have implemented the relevant 
features of the Minimal UED model in the {\tt CompHEP} 
event generator \cite{Pukhov:1999gg}.
The Minimal Supersymmetric Standard Model (MSSM)
is already available in {\tt CompHEP} since version~41.10. 
We incorporated all  $n=1$ and $n=2$ KK modes as new particles, 
with the proper interactions, widths, and one-loop corrected 
masses~\cite{Cheng:2002iz}.
Similar to the SM case, the neutral gauge bosons at level~1, 
$Z_1$ and $\gamma_1$, are mixtures of the KK modes of the
hypercharge gauge boson and the neutral $SU(2)_W$ gauge boson.
However, as shown in~\cite{Cheng:2002iz}, the radiatively corrected
Weinberg angle at level~1 and higher is very small.
For example, $\gamma_1$, which is the LKP in the minimal
UED model, is mostly the KK mode of the hypercharge gauge boson.
For simplicity, in the code we neglected neutral gauge boson mixing
for $n\ge 1$.

\section{Collider Search for Level 2 KK Particles}
\label{sec:level2}

In this section we shall consider the prospects for discovery of 
level 2 Kaluza-Klein particles in UED. Our notation and conventions follow 
those of Ref.~\cite{Cheng:2002ab}. For example, $SU(2)_W$-doublet
($SU(2)_W$-singlet) KK fermions are denoted by capital (lowercase) letters.
The KK level $n$ is denoted by a subscript.

\subsection{Phenomenology of Level 2 Fermions}
\label{sec:f2}

We begin our discussion with the $n=2$ KK fermions.
Since the KK mass spectrum is pretty degenerate,
the production cross-sections at the LHC are mostly determined 
by the strength of the KK particle interactions with
the proton constituents. As KK quarks carry color, 
we expect their production rates to be much higher than
those of KK leptons. We shall therefore concentrate on the 
case of KK quarks only.

\begin{figure}[t]
\includegraphics[width=8cm]{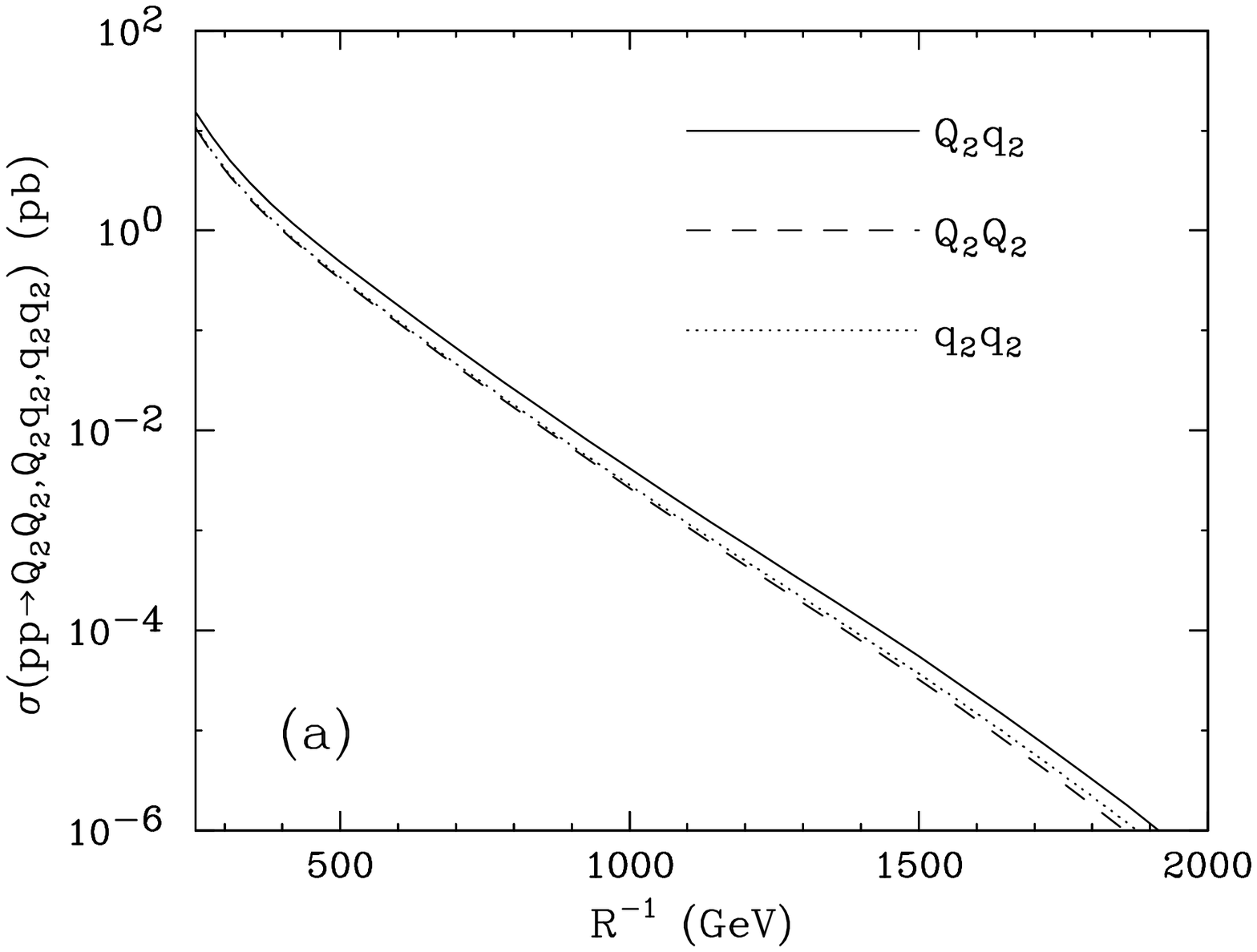}
\includegraphics[width=8cm]{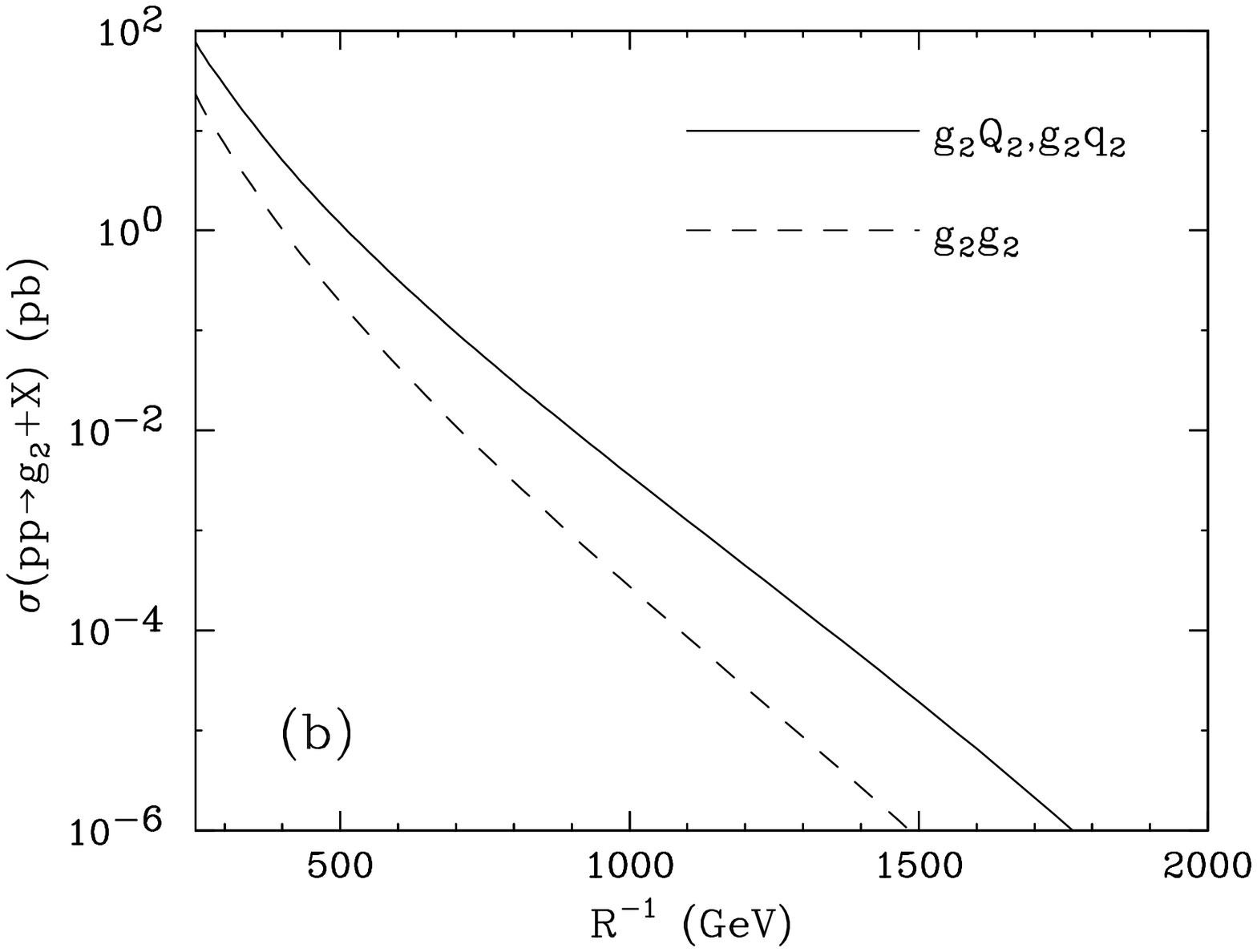}
\caption{Strong production of $n=2$ KK particles at the LHC:
(a) KK-quark pair production; 
(b) KK-quark/KK-gluon associated production and KK-gluon pair production.
The cross-sections have been summed over all quark flavors and
also include charge-conjugated contributions such as $Q_2\bar{q}_2$,
$\bar{Q}_2q_2$, $g_2\bar{Q}_2$, etc.
\label{fig:sigma_q2g2}}
\end{figure}

In principle, there are two mechanisms for producing $n=2$ KK quarks 
at the LHC: through KK-number conserving interactions,
or through KK-number violating (but KK-parity conserving)
interactions. The KK number conserving QCD interactions allow
production of KK quarks either in pairs or singly (in association with 
the $n=2$ KK mode of a gauge boson). The corresponding production 
cross-sections are shown in Fig.~\ref{fig:sigma_q2g2}
(the cross-sections for producing $n=1$ KK quarks have been 
calculated in~\cite{Rizzo:2001sd,Macesanu:2002db,Smillie:2005ar}). 
In Fig.~\ref{fig:sigma_q2g2}a we show the cross-sections (in pb)
for $n=2$ KK-quark pair production, 
while in Fig.~\ref{fig:sigma_q2g2}b we show the 
results for $n=2$ KK-quark/KK-gluon associated production 
and for $n=2$ KK-gluon pair production. We plot the results 
versus $R^{-1}$, and one should keep in mind that the masses
of the $n=2$ particles are roughly $2/R$. In calculating the 
cross-sections of Fig.~\ref{fig:sigma_q2g2} we consider 5 
partonic quark flavors in the proton along with the gluon.
We sum over the final state quark flavors and include
charge-conjugated contributions. We used CTEQ5L parton distributions 
\cite{Lai:1999wy} and choose the scale of the strong coupling constant
$\alpha_s$ to be equal to the parton level center of mass energy.
All calculations are done with {\tt CompHEP}~\cite{Pukhov:1999gg}
with our implementation of the Minimal UED model.

Several comments are in order. First, Fig.~\ref{fig:sigma_q2g2}
displays a severe kinematic suppression of the cross-sections
at large KK masses. This is familiar from the case of SUSY, where 
the ultimate LHC reach for colored superpartners extends only to about
3 TeV. Notice the different mass dependence of the cross-sections for
the three types of final states with $n=2$ particles: 
quark-quark, quark-gluon, and gluon-gluon. This can be easily understood
in terms of the structure functions of the quarks and gluon 
inside the proton. We also observe minor differences in the 
cross-sections for pair production of KK quarks with different 
$SU(2)_W$ quantum numbers. This is partially due to the different masses
for $SU(2)_W$-doublet and $SU(2)_W$-singlet quarks (see Fig.~\ref{fig:spectrum}), 
and the remaining difference is due to the contributions 
from diagrams with electroweak gauge bosons. Notice that since
the cross-sections in Fig.~\ref{fig:sigma_q2g2}a
are summed over charge conjugated final states, 
the mixed case of $Q_2q_2$ contains twice as many quark-antiquark 
contributions (compare $Q_2\bar{q}_2+\bar{Q}_2q_2$ to 
$q_2\bar{q}_2$ or $Q_2\bar{Q}_2$ alone).

If we compare the cross-sections for $n=2$ KK quark production 
to the cross-sections for producing squarks of similar masses in SUSY,
we realize that the production rates are higher in UED. 
This is due to several reasons. Consider, for example, 
$s$-channel processes. Well above 
threshold, the UED cross-sections are larger by a factor of 
4~\cite{Battaglia:2005zf}. One factor of 2 is due to the fact 
that in UED the particle content at $n\ge 1$ is duplicated --
for example, there are both left-handed and right-handed
$SU(2)_W$-doublet KK fermions, while in SUSY there are only ``left-handed''
$SU(2)_W$-doublet squarks. Another factor of 2 comes from 
the different angular distribution for fermions, $1+\cos^2\theta$,
versus scalars, $1-\cos^2\theta$. When integrated over all 
angles, this accounts for the second factor of 2 difference. 
Furthermore, at the LHC new heavy particles are produced 
close to threshold, due to the steeply falling parton luminosities.
In SUSY, the new particles (squarks) are scalars, and the threshold 
suppression of the cross-sections is $\sim \beta^3$, while in
UED the KK-quarks are fermions, and the threshold suppression 
of the cross-section is only $\beta$.
This distinct threshold behavior of the production cross-sections
further enhances the difference between SUSY and UED. For example, we find that for
$R^{-1}=500$ GeV the pair production cross-section for charm KK-quarks
is about 6 times larger than the cross-section for charm squarks.
For processes involving first generation KK-quarks, 
where $t$-channel diagrams contribute significantly, the
effect can be even bigger. For example, up KK-quark production
and up squark production differ by about factor of 8.

In Fig.~\ref{fig:sigma_q2g2} we have only considered 
production due to KK number conserving bulk interactions.
The main advantage of those processes is that the corresponding
couplings are unsuppressed. However, the disadvantage is that
we need to produce {\em two} heavy particles, each of mass $\sim 2/R$,
which leads to a kinematic suppression. In order to overcome this
problem, one could in principle consider the single production 
of $n=2$ KK quarks through KK number violating, but KK parity 
conserving interactions, for example
\begin{equation}
{\bar Q}_2 \gamma^\mu T^a P_L Q_0 {A^a_0}_\mu\ ,
\label{Q2Q0A0}
\end{equation}
where $A_\mu^a$ is a SM gauge field and $T^a$ is the corresponding 
group generator. However, (\ref{Q2Q0A0}) is forbidden by gauge invariance,
and the lowest order coupling of an $n=2$ KK quark to two SM particles 
has the form \cite{Cheng:2002iz}
\begin{equation}
{\bar Q}_2 \sigma^{\mu\nu} T^a P_L Q_0 {F^a_0}_{\mu\nu}\ .
\label{Q2Q0A0HD}
\end{equation}
Such operators may in principle be present, as they may be
generated at the scale $\Lambda$ by the unknown physics at higher scales.
However, being higher dimensional, we expect them to be suppressed at least
by $1/\Lambda$, hence in our subsequent analysis we shall neglect them.

\begin{figure}[t]
\includegraphics[width=8.0cm]{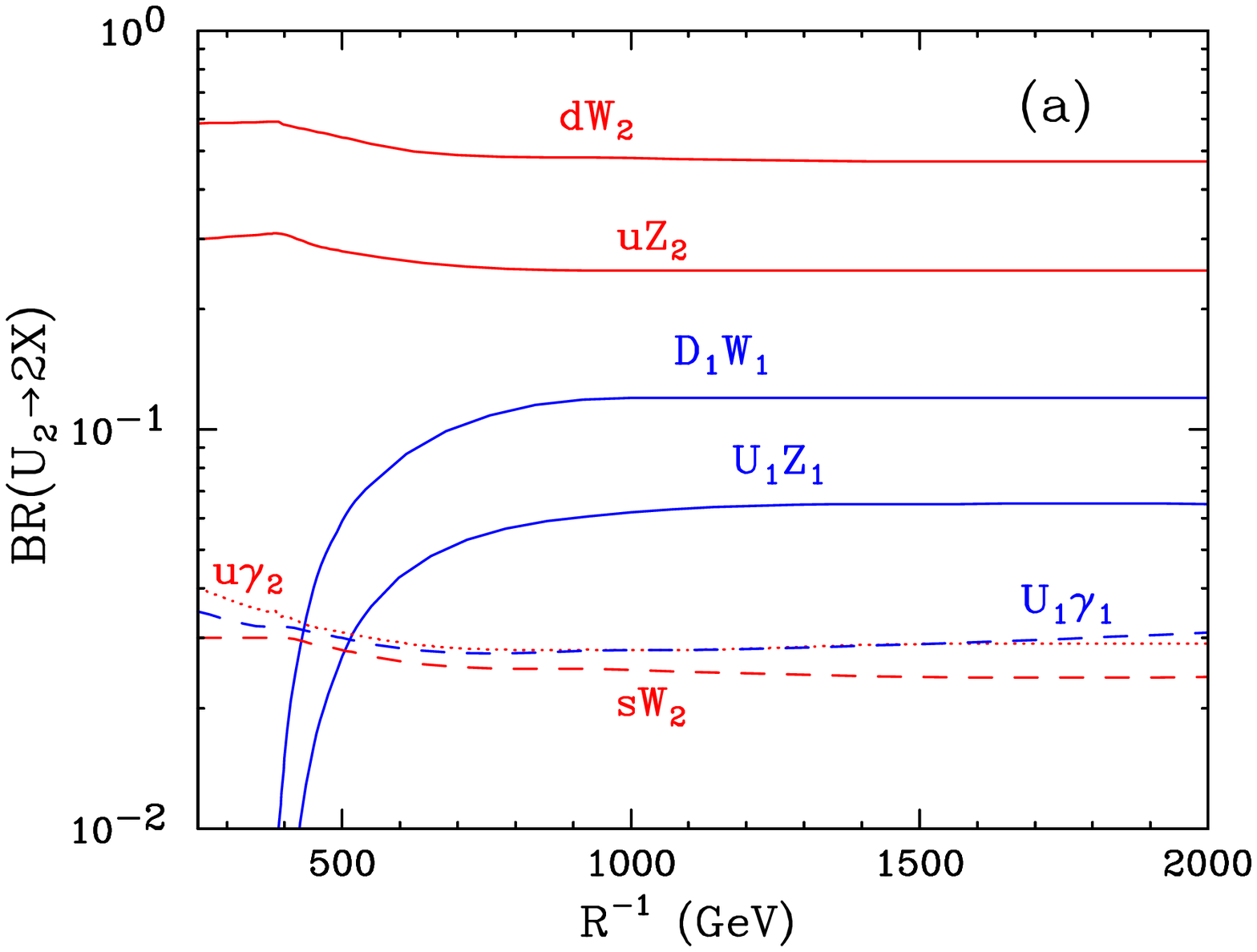}
\includegraphics[width=8.0cm]{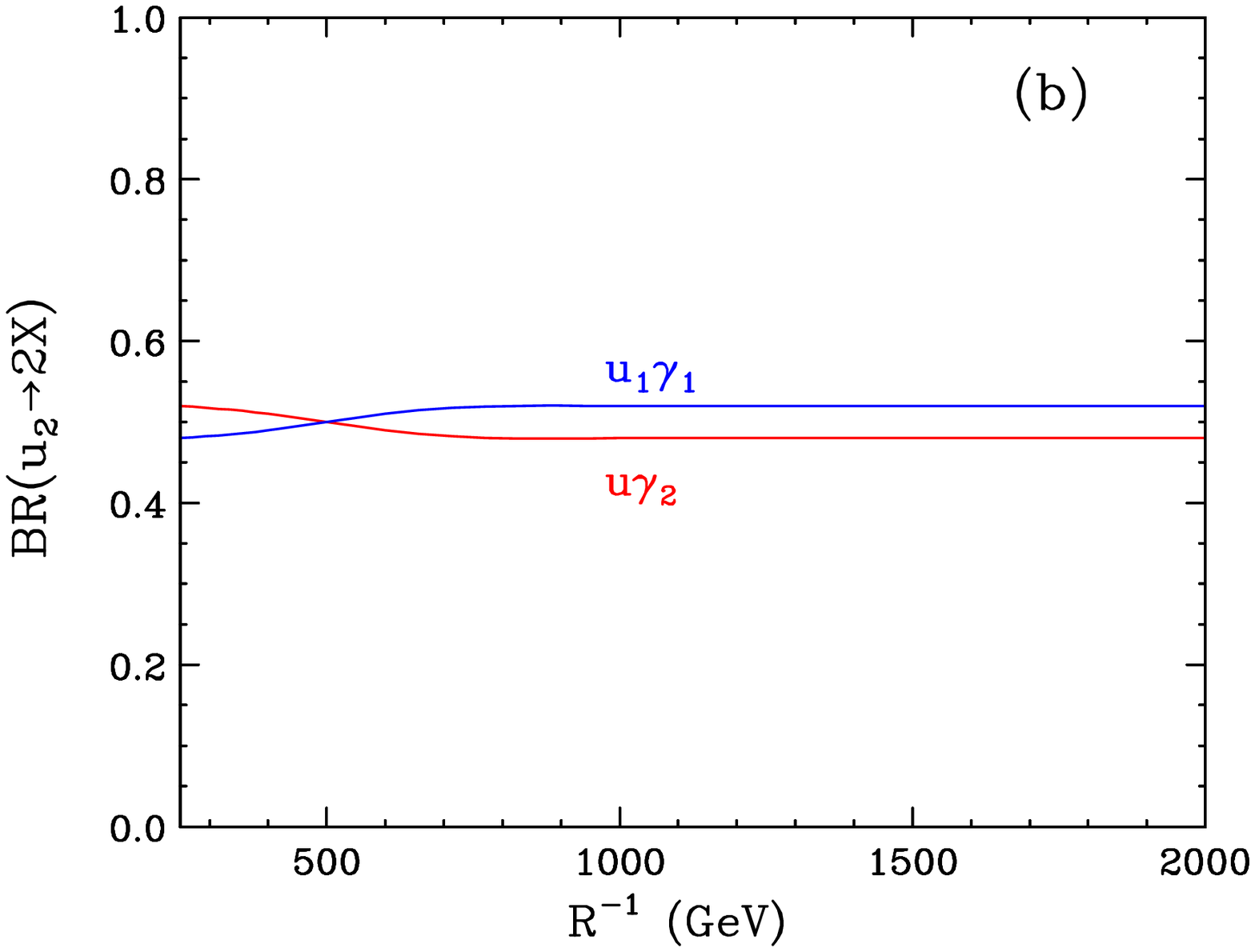}
\caption{Branching fractions of the level 2 ``up'' quarks
versus $R^{-1}$; for (a) the $SU(2)_W$-doublet quark $U_2$ and 
(b) the $SU(2)_W$-singlet quark $u_2$.
\label{fig:br_q2}}
\end{figure}

Having determined the production rates of level 2 KK quarks, we 
now turn to the discussion of their experimental signatures.
To this end we need to determine the possible decay modes of 
$Q_2$ and $q_2$. At each level $n$, the KK quarks are among 
the heaviest states in the KK spectrum and can decay promptly to lighter 
KK modes (this is true for the top KK 
modes~\cite{Carone:2003ms,DePree:2005yv} as well). 
As can be seen from Fig.~\ref{fig:spectrum}, 
the KK gluon is always heavier than the KK quarks, 
so the two body decays of KK quarks to KK gluons are closed. Instead, 
$n=2$ KK quarks will decay to the KK modes of the electroweak 
gauge bosons which are lighter. The branching fractions for 
$n=2$ ``up''-type KK quarks are shown in Fig.~\ref{fig:br_q2}.
Fig.~\ref{fig:br_q2}a (Fig.~\ref{fig:br_q2}b) is for the case of the
$SU(2)_W$-doublet quark $U_2$ (the $SU(2)_W$-singlet quark $u_2$).
The results for the ``down''-type KK quarks are similar.
We observe in Fig.~\ref{fig:br_q2} that the branching fractions
are almost independent of $R^{-1}$, unless one is close to threshold.
This feature will persist for all branching ratios of KK particles 
which will be shown later. 

Once we ignore the KK number violating coupling (\ref{Q2Q0A0HD}),
only decays which conserve the total KK number $n$ are allowed.
The case of the $SU(2)_W$-singlet quarks such as $u_2$ is simpler, 
since they only couple to the hypercharge gauge bosons. Recall that 
at $n\ge1$ the hypercharge component is almost entirely contained in
the $\gamma$ KK mode~\cite{Cheng:2002iz}. We therefore expect a singlet 
KK quark $q_2$ to decay to either $q_1\gamma_1$ or $q_0\gamma_2$, as
seen in Fig.~\ref{fig:br_q2}b. The case of an $SU(2)_W$-doublet quark
$Q_2$ is much more complicated, since $Q_2$ couples to the 
(KK modes of the) weak gauge bosons as well, and many more 
two-body final states are possible. Since the weak coupling is 
larger than the hypercharge coupling, the decays to $W$ and $Z$ 
KK modes dominate, with 
$BR(Q_2\to Q'_0W_2)/BR(Q_2\to Q_0Z_2)=2$ and
$BR(Q_2\to Q'_1W_1)/BR(Q_2\to Q_1Z_1)=2$, as evidenced in
Fig.~\ref{fig:br_q2}a. The branching fractions to the 
$\gamma$ KK modes are only on the order of a few percent.
The threshold behavior seen in Fig.~\ref{fig:br_q2}a
near $R^{-1}=400$ GeV is due to the finite masses for the 
SM $W$ and $Z$ bosons, which enter the tree-level masses of 
$W^\pm_1$ and $Z_1$. Since the mass splitting of the KK modes
is due to the radiative corrections, which are proportional 
to $R^{-1}$, the channels with $W^\pm_1$ and $Z_1$ open up 
only for sufficiently large $R^{-1}$.


We are now in a position to discuss the experimental 
signatures of $n=2$ KK quarks. The decays to level 2 gauge bosons 
will simply contribute to the inclusive production of $\gamma_2$, 
$Z_2$ and $W^\pm_2$, which will be discussed at length 
later in Section~\ref{sec:V2}. On the other hand, the decays 
to two $n=1$ KK modes will contribute to the inclusive production 
of $n=1$ KK particles which was discussed in \cite{Cheng:2002ab}.
Naturally, the direct pair production of the lighter $n=1$ KK modes 
has a much larger cross-section. Therefore, the indirect production of 
$n=1$ KK modes from the decays of $n=2$ particles can be easily 
swamped by the direct $n=1$ signals and the SM backgrounds. 
For example, the experimental
signature for an $n=2$ KK quark decaying as $Q_2\to Q_1\gamma_1$ 
($q_2\to q_1\gamma_1$) is indistinguishable from a single $Q_1$ 
($q_1$). This is because $\gamma_1$ does not interact within 
the detector, and there are at least two additional $\gamma_1$ 
particles in each event, so that we cannot determine how many 
$\gamma_1$ particles caused the measured amount of missing energy.
The decays to $W_1$ and $Z_1$ may, however, lead to
final states with up to four $n=1$ particles, each with a leptonic
decay mode. The resulting multilepton signatures $N\ell+\met$ with $N\ge5$
are therefore very clean and potentially observable.
Distinguishing those events from direct $n=1$ pair 
production would be an important step in establishing the presence of the 
$n=2$ level of the quark KK tower. Unfortunately, the $n=2$ 
sample is statistically very limited and this analysis 
appears very challenging. We postpone it for future 
work~\cite{DKM}.

\begin{figure}[t]
\includegraphics[width=8.0cm]{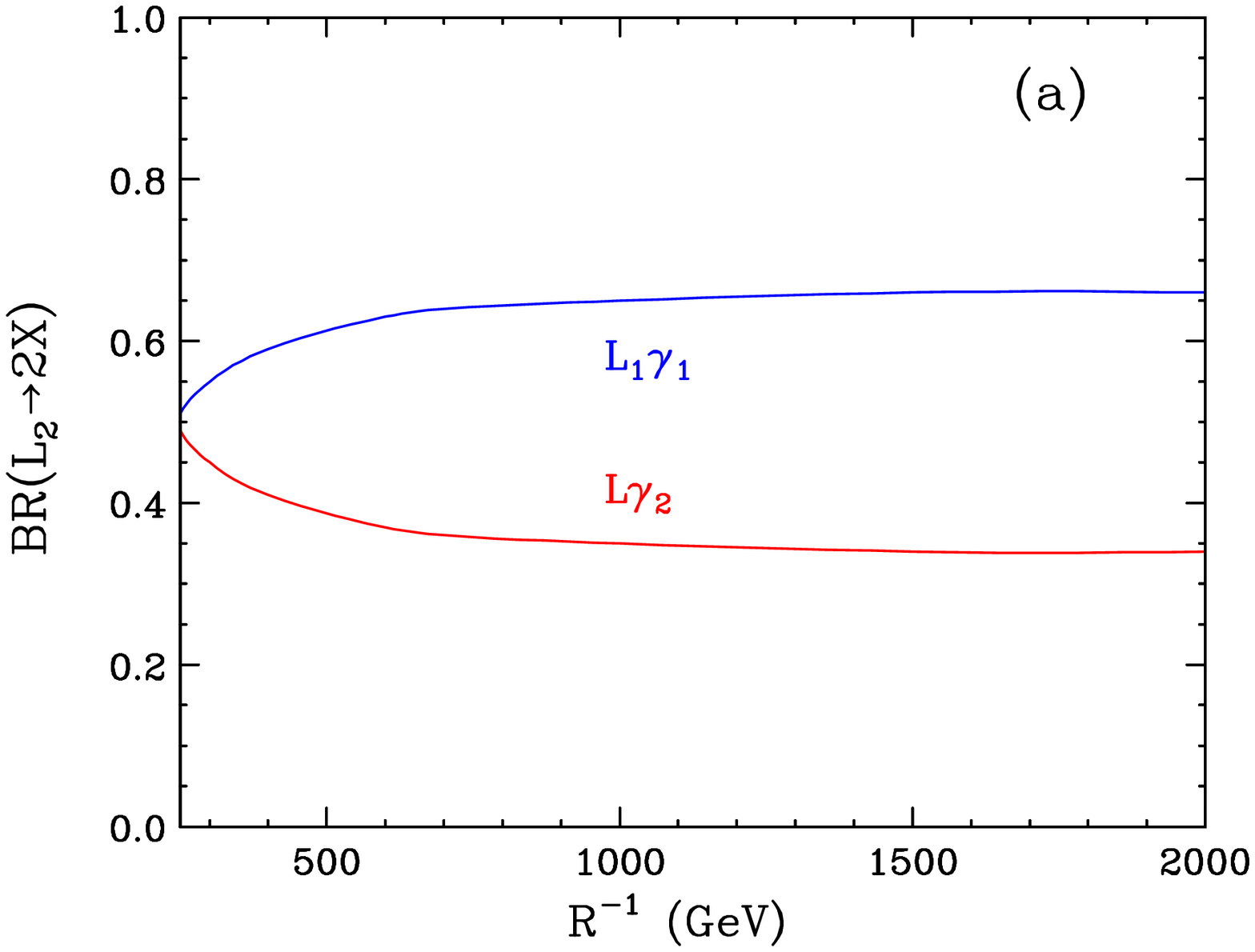}
\includegraphics[width=8.0cm]{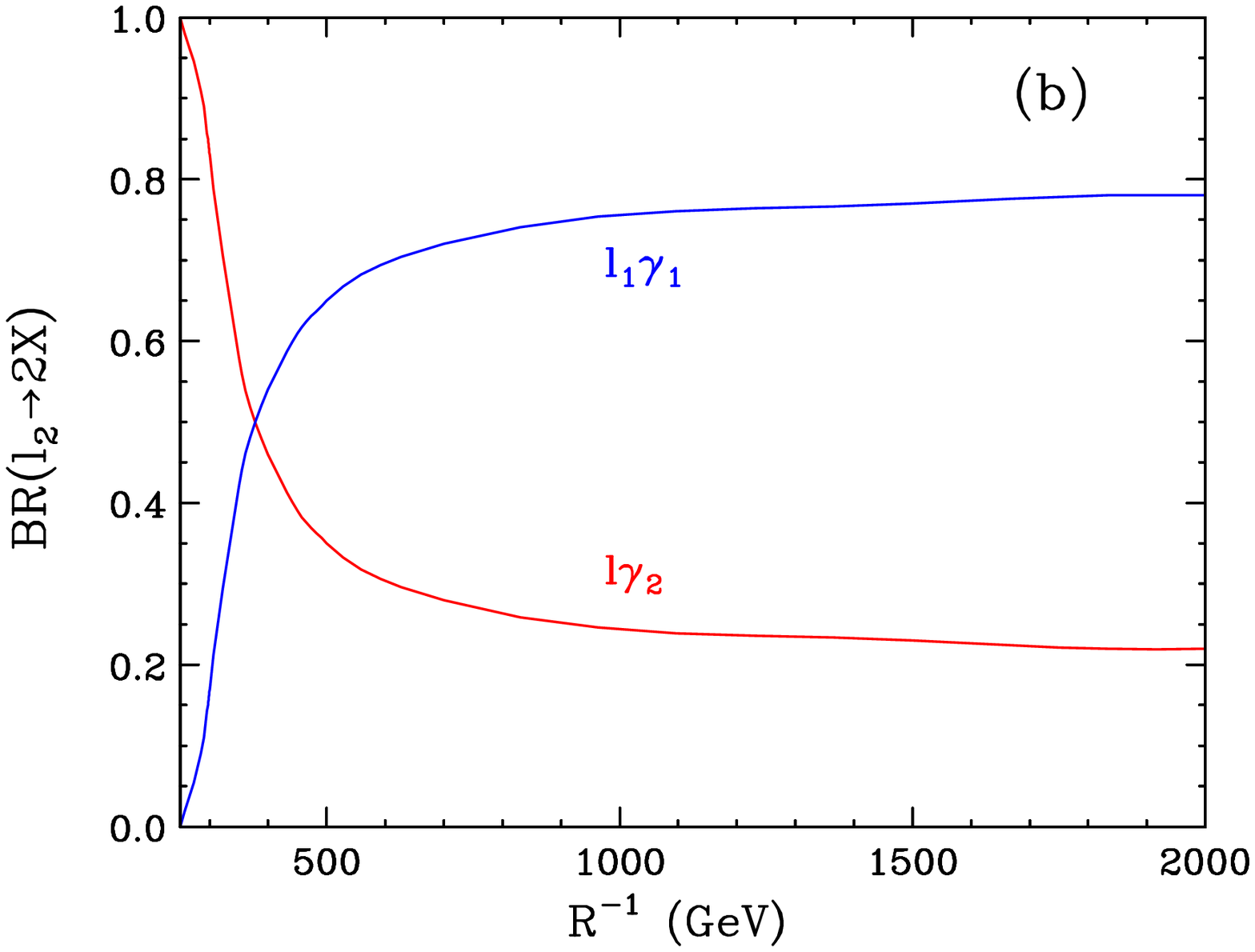}
\caption{The same as Fig.~\ref{fig:br_q2}
but for the level 2 KK electrons:
(a) the $SU(2)_W$-doublet $E_2$ and 
(b) the $SU(2)_W$-singlet $e_2$.
\label{fig:br_l2}}
\end{figure}

Much of the previous discussion applies directly to the 
level 2 KK leptons. Assuming the absence of the 
KK number violating coupling analogous to (\ref{Q2Q0A0HD}),
the branching fractions of the $n=2$ KK electrons 
are shown in Fig.~\ref{fig:br_l2}. At each KK level, the
KK modes of the weak gauge bosons are heavier than the
KK leptons, therefore the only allowed decays are to $\gamma_2$ 
and $\gamma_1$. Just like KK quarks, KK leptons can be produced directly, 
through KK number conserving couplings, or indirectly, in $W^\pm_2$
and $Z_2$ decays. In either case, the resulting cross-sections are 
too small to be of interest at the LHC.

\subsection{Level 2 Gauge Bosons}
\label{sec:V2}

We now discuss the collider phenomenology of the $n=2$ gauge bosons $V_2$.
As we shall see, the KK gauge bosons offer the best prospects for 
detecting the $n=2$ structure, since they have direct (but not tree level)
couplings to SM particles, and can be discovered as resonances, e.g. in the dijet 
or dilepton channels. This is in contrast to the case of $n=2$ KK fermions, 
which, under the assumptions of Sec.~\ref{sec:f2}, do not have fully visible 
decay modes. Bump hunting will also help discriminate between $n=2$ and
$n=1$ KK particles, since the latter are KK-parity odd, and necessarily 
decay to the invisible $\gamma_1$.

\begin{figure}[t]
\includegraphics[width=8cm]{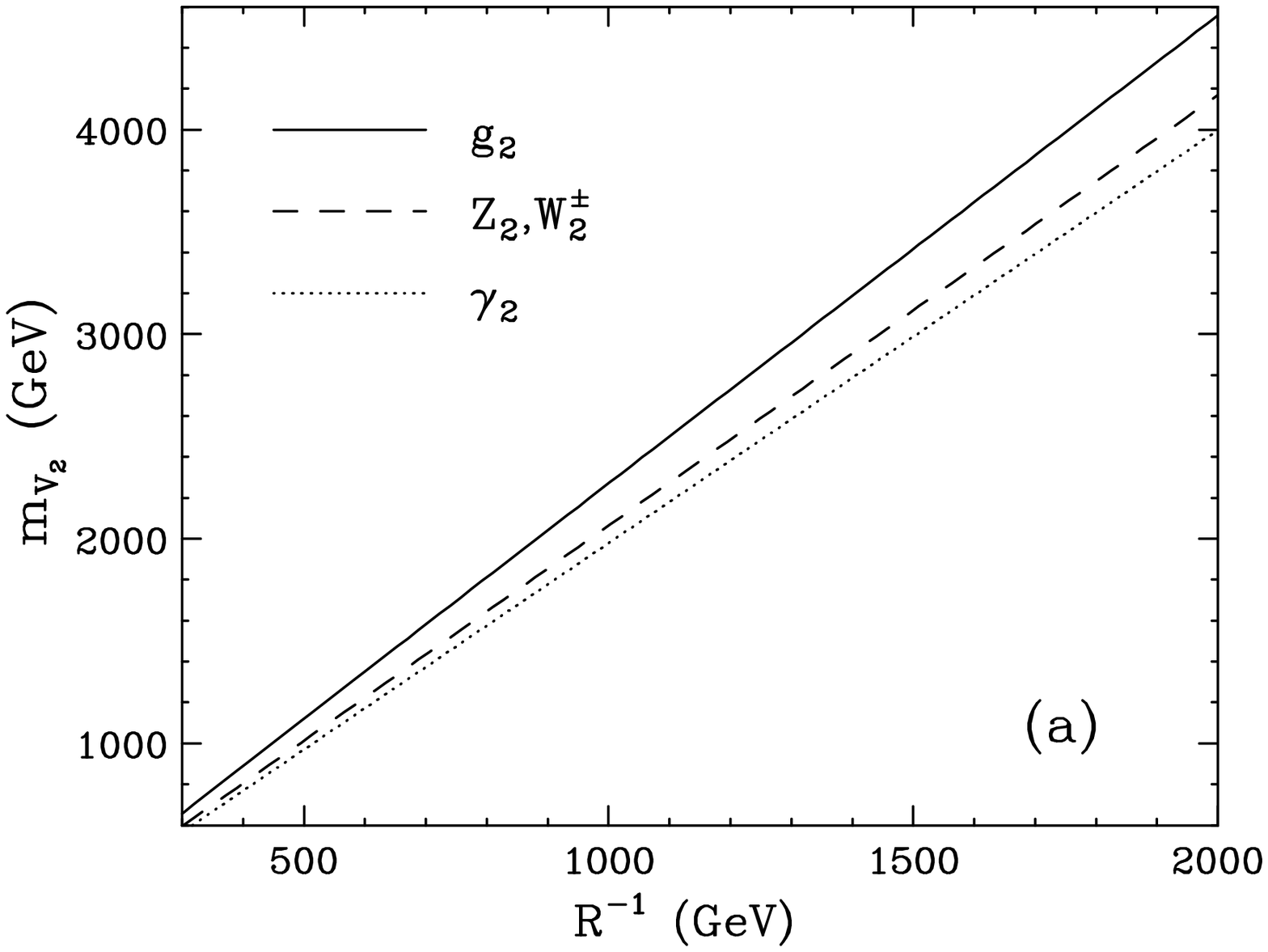}
\includegraphics[width=8cm]{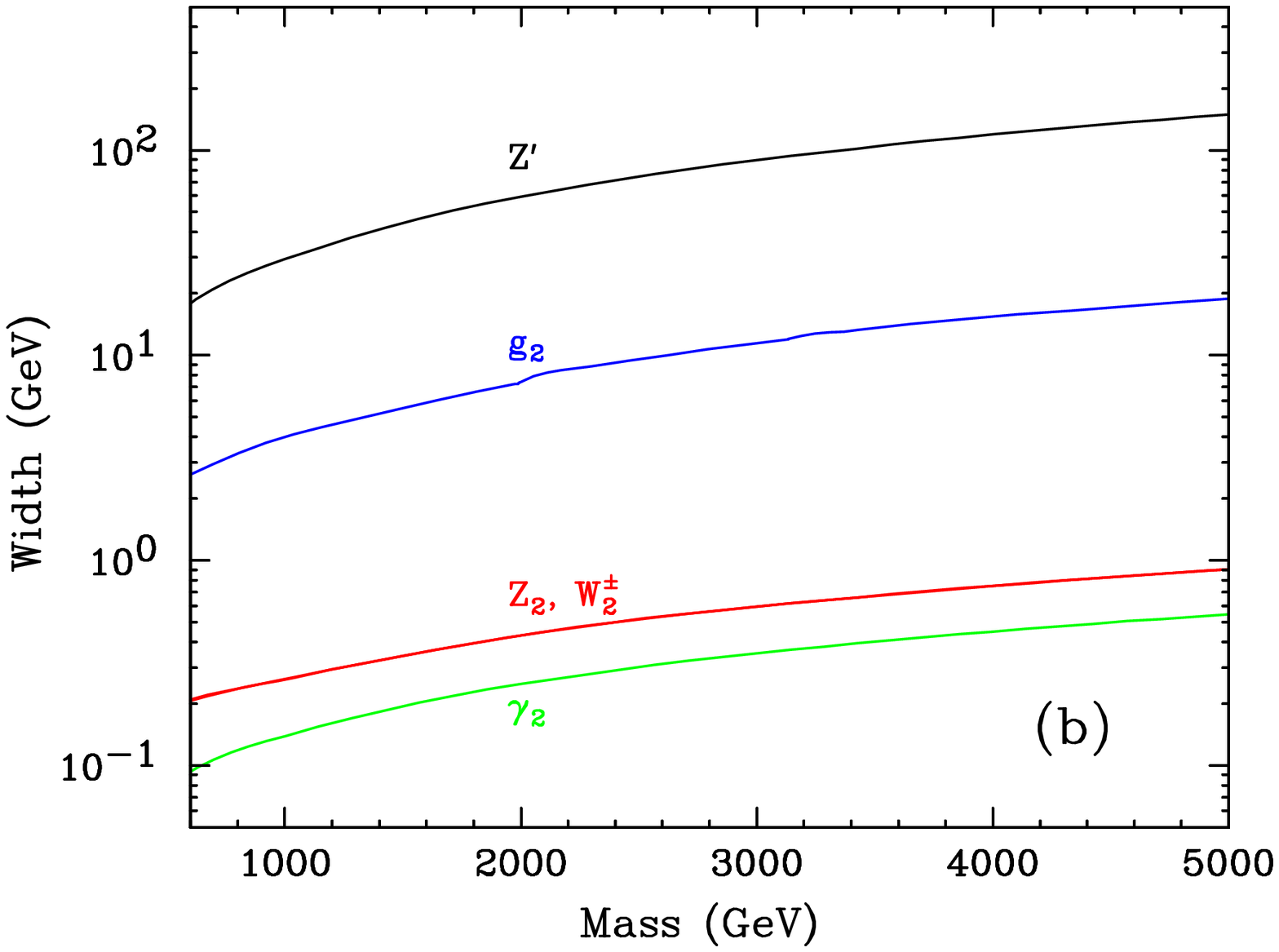}
\caption{(a) Masses of the four $n=2$ KK gauge bosons
as a function of $R^{-1}$. (b) Total widths of the $n=2$ KK 
gauge bosons as a function of the corresponding mass.
We also show the width of a generic $Z'$ whose couplings to the 
SM particles are the same as those of the $Z$-boson.
\label{fig:Gammas}}
\end{figure}

There are four $n=2$ KK gauge bosons: the KK ``photon'' $\gamma_2$, 
the KK ``$Z$-boson'' $Z_2$, the KK ``$W$-boson'' $W^\pm_2$,
and the KK gluon $g_2$. Recall that the Weinberg angle at $n=2$ is 
very small, so that $\gamma_2$ is mostly the KK mode of the 
hypercharge gauge boson and $Z_2$ is mostly the KK mode of the 
neutral $W$-boson of the SM. An important consequence of the
extra dimensional nature of the model is that all four of the 
$n=2$ KK gauge bosons are relatively degenerate, as shown in 
Fig.~\ref{fig:Gammas}a. The masses are all roughly equal to $2/R$. 
The mass splitting between the KK gauge bosons is almost entirely 
due to radiative corrections, which in the Minimal UED model
yield the mass hierarchy $m_{g_2}>m_{W_2}\sim m_{Z_2}>m_{\gamma_2}$.
The KK gluon receives the largest corrections and is the 
heaviest particle in the KK spectrum at each level $n$.
The $W^\pm_2$ and $Z_2$ particles are degenerate to a very high degree, 
due to $SU(2)_W$ symmetry. 

The KK number conserving interactions allow an $n=2$ KK gauge boson $V_2$
to decay to two $n=1$ particles, or to one $n=2$ KK particle and one
$n=0$ (i.e., Standard Model) particle, provided that the decays are allowed 
by phase space. For example, the partial widths to fermion final states are
given by
\begin{eqnarray}
\Gamma( V_2 \rightarrow f_2 \bar{f}_0) &=& \frac{c^2 g^2}{48\pi m^3_{V_2}} 
\left ( m^2_{V_2}-m^2_{f_2}-m^2_{f_0} + \frac{m^4_{V_2} - (m^2_{f_2}-m^2_{f_0})^2}{m^2_{V_2}} \right ) \nonumber \\
&&\times \sqrt{\left ( m^2_{V_2} - (m_{f_2}-m_{f_0})^2 \right ) 
                           \left ( m^2_{V_2} - (m_{f_2}+m_{f_0})^2 \right )} \label{G220} \\
&\approx& \frac{c^2 g^2}{48\pi m^3_{V_2}} 
\left ( m^2_{V_2}-m^2_{f_2}  \right )^2 \left ( 1+\frac{m^2_{V_2}+m^2_{f_2}}{m^2_{V_2}} \right ) \nonumber \\
&\approx& \frac{{c^2 g^2} m_{V_2}}{4\pi} 
\left ( \frac{\hat{\delta}m_{V_2}}{m_2} - \frac{\hat{\delta}m_{f_2}}{m_2} \right )^2\ , \nonumber 
\end{eqnarray}
\begin{eqnarray}
\Gamma( V_2 \rightarrow f_1 \bar{f}_1) &=& \frac{c^2 g^2}{24\pi m^2_{V_2}} 
\left ( m^2_{V_2}-4m^2_{f_1} \right )^\frac{3}{2} \label{G211}\\
&\approx& \frac{{c^2 g^2} m_{V_2}}{6\sqrt{2}\pi} 
\left ( \frac{\hat{\delta}m_{V_2}}{m_2} - \frac{\hat{\delta}m_{f_1}}{m_1} \right )^\frac{3}{2} \left ( \frac{m_2}{m_{V_2}} \right )^3 \nonumber \\
&\approx& \frac{{c^2 g^2} m_{V_2}}{6\sqrt{2}\pi} 
\left ( \frac{\hat{\delta}m_{V_2}}{m_2} - \frac{\hat{\delta}m_{f_1}}{m_1} \right )^\frac{3}{2}\ , \nonumber
\end{eqnarray}
where $c\approx YN_c^f/2$ for $V_2\approx\gamma_2$, $c\approx N_c^f/2$ for $V_2\approx Z_2$, 
$c=V_{CKM}N_c^f/\sqrt{2}$ for $V_2=W^\pm_2$ and $c=1/\sqrt{2}$ for $V_2=g_2$,
with $Y$ being the fermion hypercharge in the normalization
$Q=T_3+Y/2$, $V_{CKM}$ is the CKM mixing matrix,
and $N_c^f=3$ for $f=q$ and $N_c^f=1$ for $f=\ell$.
Here $\hat{\delta}m$ stands for the total radiative correction to a KK mass
$m$, including both bulk and boundary contributions~\cite{Cheng:2002iz},
$m_2\equiv 2/R$, and $g$ is the corresponding gauge coupling.
The first lines in (\ref{G220}) and (\ref{G211}) give the exact result, while the
last lines are the approximate formulas derived in~\cite{Cheng:2002ab}
as leading order expansions in $\hat{\delta}m/m$.
The second line in (\ref{G220}) is an approximation neglecting the 
SM fermion mass $m_{f_0}$. The second line in (\ref{G211}) is 
an alternative approximation which incorporates subleading but numerically 
important terms. In our code we have programmed the exact expressions and 
quote the approximations here only for completeness.

Note that the KK number conserving decays of the $n=2$ KK gauge 
bosons are suppressed by phase space. This is evident from the
approximate expressions in eqs.~(\ref{G220}) and (\ref{G211}).
The partial widths are proportional to the one-loop corrections, 
which open up the available phase space and allow the corresponding 
decay mode to take place. However, not all of the fermionic final states
are available, for example, $Z_2$ and $W^\pm_2$ have no hadronic decay modes
to level 1 or 2, while $\gamma_2$ has {\em no} KK number conserving 
decay modes at all.

The $n=2$ KK gauge bosons also have KK number violating couplings 
which can be generated either radiatively from bulk interactions, 
or directly at the scale $\Lambda$~\cite{Cheng:2002iz}. 
For example, the operator
\begin{equation}
{\bar f}_0 \gamma^\mu T^a P_L f_0 {A^a_2}_\mu\ 
\label{Q0Q0A2}
\end{equation}
couples $V_2$ directly to SM fermions $f_0$, and leads to the
the following $V_2$ partial width
\begin{eqnarray}
\Gamma( V_2 \rightarrow f_0 \bar{f}_0) &=& \frac{c^2 {g^2} m_{V_2}}{12\pi} 
\left ( \frac{\bar{\delta}m_{V_2}}{m_2} - \frac{\bar{\delta}m_{f_2}}{m_2} \right )^2
\left ( 1 - \frac{m^2_{f_0}}{m^2_{V_2}} \right ) 
\sqrt{\left ( 1 - 4\frac{m^2_{f_0}}{m^2_{V_2}} \right )} \label{G200}\\
&\approx& \frac{c^2 {g^2} m_{V_2}}{12\pi} 
\left ( \frac{\bar{\delta}m_{V_2}}{m_2} - \frac{\bar{\delta}m_{f_2}}{m_2} \right )^2\ , \nonumber
\end{eqnarray}
where $\bar{\delta}m$ stands for a mass correction due to boundary terms only
\cite{Cheng:2002iz}. In the second line we have neglected the SM fermion mass
$m_{f_0}$, recovering the result from \cite{Cheng:2002ab}.

As we see from (\ref{G200}), the KK number violating decay is also suppressed, 
this time by a loop factor, and is proportional to the size of the 
radiative corrections to the corresponding KK masses. In spite of 
this suppression, the $V_2\to f_0\bar{f}_0$ decays is most promising
for experimental discovery. As long as the final state fermions
can be reconstructed, the $V_2$ particle can be looked for
as a bump in the invariant mass distribution of its decay products.
In this sense, the search is very similar to $Z'$ searches, with one 
major difference. Since {\em all} partial widths (\ref{G220}-\ref{G200}) 
are suppressed, the {\em total} width of $V_2$ is much smaller than 
the width of a typical $Z'$. This is illustrated in Fig.~\ref{fig:Gammas}b,
where we plot the widths of the KK particles $\gamma_2$,
$W^\pm_2$, $Z_2$ and $g_2$ in UED, as a function of the corresponding particle mass,
and contrast to the width of a $Z'$ with SM-like couplings.
We see that the widths of the KK gauge bosons are extremely small.
This has important ramifications for the experimental search, 
since the width of the resonance will then be determined by the 
experimental resolution, rather than the intrinsic particle width.
In this sense the width must be included in the set of basic parameters 
of a $Z'$ search \cite{Carena:2004xs}.

\begin{figure}[t]
\includegraphics[width=10cm]{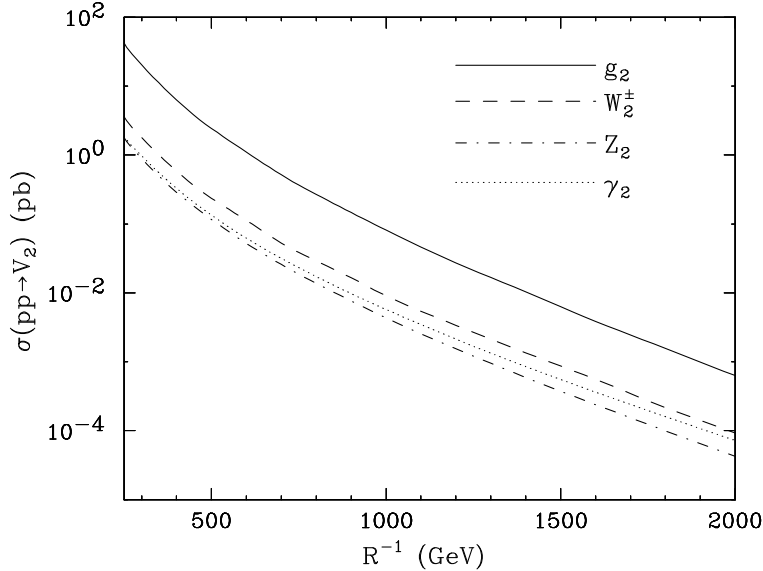}
\caption{Cross-sections for single production of 
level 2 KK gauge bosons through the KK number violating couplings
(\ref{Q0Q0A2}).
\label{fig:sigma_V2}}
\end{figure}

Before we elaborate on the experimental signatures of the $n=2$ KK gauge bosons, 
let us briefly discuss their production. There are three basic mechanisms:

{\bf 1. Single production through the KK number violating operator (\ref{Q0Q0A2}).}
The corresponding cross-sections are shown in Fig.~\ref{fig:sigma_V2} 
as a function of $R^{-1}$. One might expect that these processes will be important, 
especially at large masses, since we need to make only a single 
heavy $n=2$ particle, alleviating the kinematic suppression.
If we compare the mass dependence of the Drell-Yan cross-sections 
in Fig.~\ref{fig:sigma_V2} to the mass dependence of the $n=2$ pair production
cross-sections from Fig.~\ref{fig:sigma_q2g2}, indeed we see that the former drop less 
steeply with $R^{-1}$ and become dominant at large $R^{-1}$.
On the other hand, the Drell-Yan processes of Fig.~\ref{fig:sigma_V2}
are mediated by a KK number violating operator (\ref{Q0Q0A2})
and the coupling of a $V_2$ to SM particles is radiatively suppressed. 
This is another crucial difference with the case of a generic $Z'$,
whose couplings typically have the size of a normal gauge coupling
and are unsuppressed~\cite{Carena:2004xs}. 

Notice the roughly similar size of the
four cross-sections shown in Fig.~\ref{fig:sigma_V2}. This is somewhat
surprising, since the cross-sections scale as the corresponding 
gauge coupling squared, and one would have expected a wider spread 
in the values of the four cross-sections. This is due to a couple of things.
First, for a given $R^{-1}$, the masses of the four $n=2$ KK gauge bosons 
are different, with $m_{g_2}>m_{W_2}\sim m_{Z_2}>m_{\gamma_2}$. 
Therefore, for a given $R^{-1}$, the heavier particles suffer a suppression.
This explains to an extent why the cross-section for $\gamma_2$ is 
not the smallest of the four, and why the cross-section for $g_2$ is
not as large as one would expect. There is, however, a second effect, 
which goes in the same direction. The coupling (\ref{Q0Q0A2})
is also proportional to the mass corrections of the corresponding particles:
\begin{equation}
\frac{\bar\delta m_{V_2}}{m_{V_2}} - \frac{\bar\delta m_{f_2}}{m_{f_2}}\ .
\label{gQ0Q0A2}
\end{equation}
Since the QCD corrections are the largest, for $V_2=\{\gamma_2,Z_2,W^\pm_2\}$,
the second term dominates. However, for $V_2=g_2$, the first term is actually
larger, and there is a cancellation, which further reduces the direct
KK gluon couplings to quarks.

{\bf 2. Indirect production.} The electroweak KK modes $\gamma_2$, $Z_2$ 
and $W^\pm_2$ can be produced in the decays of heavier $n=2$ particles 
such as the KK quarks and/or KK gluon.
This is well known from the case of SUSY, where the dominant production of
electroweak superpartners is often indirect -- from squark and gluino decay
chains. The indirect production rates of $\gamma_2$, $Z_2$ and $W^\pm_2$ 
due to QCD processes can be readily estimated from Figs.~\ref{fig:sigma_q2g2}
and \ref{fig:br_q2}. Notice that $BR(Q_2\to W^\pm_2)$, $BR(Q_2\to Z_2)$ and 
$BR(q_2\to \gamma_2)$ are among the largest branching fractions of the $n=2$ 
KK quarks, and we expect indirect production from QCD to be a significant 
source of electroweak $n=2$ KK modes.

{\bf 3. Direct pair production.} The $n=2$ KK modes can also be produced 
directly in pairs, through KK number conserving interactions. These processes, 
however, are kinematically suppressed, since we have to make {\em two} 
heavy particles in the final state. One would therefore expect that
they will be the least relevant source of $n=2$ KK gauge bosons.
The only exception is KK gluon pair production which is important 
and is shown in Fig.~\ref{fig:sigma_q2g2}b. We see that it is 
comparable in size to KK quark pair production and $q_2g_2$/$Q_2g_2$ 
associated production. We have also calculated the pair production 
cross-sections for the electroweak $n=2$ KK gauge bosons and confirmed 
that they are very small, hence we shall neglect them in our analysis below.

\begin{figure}[t]
\includegraphics[width=8cm]{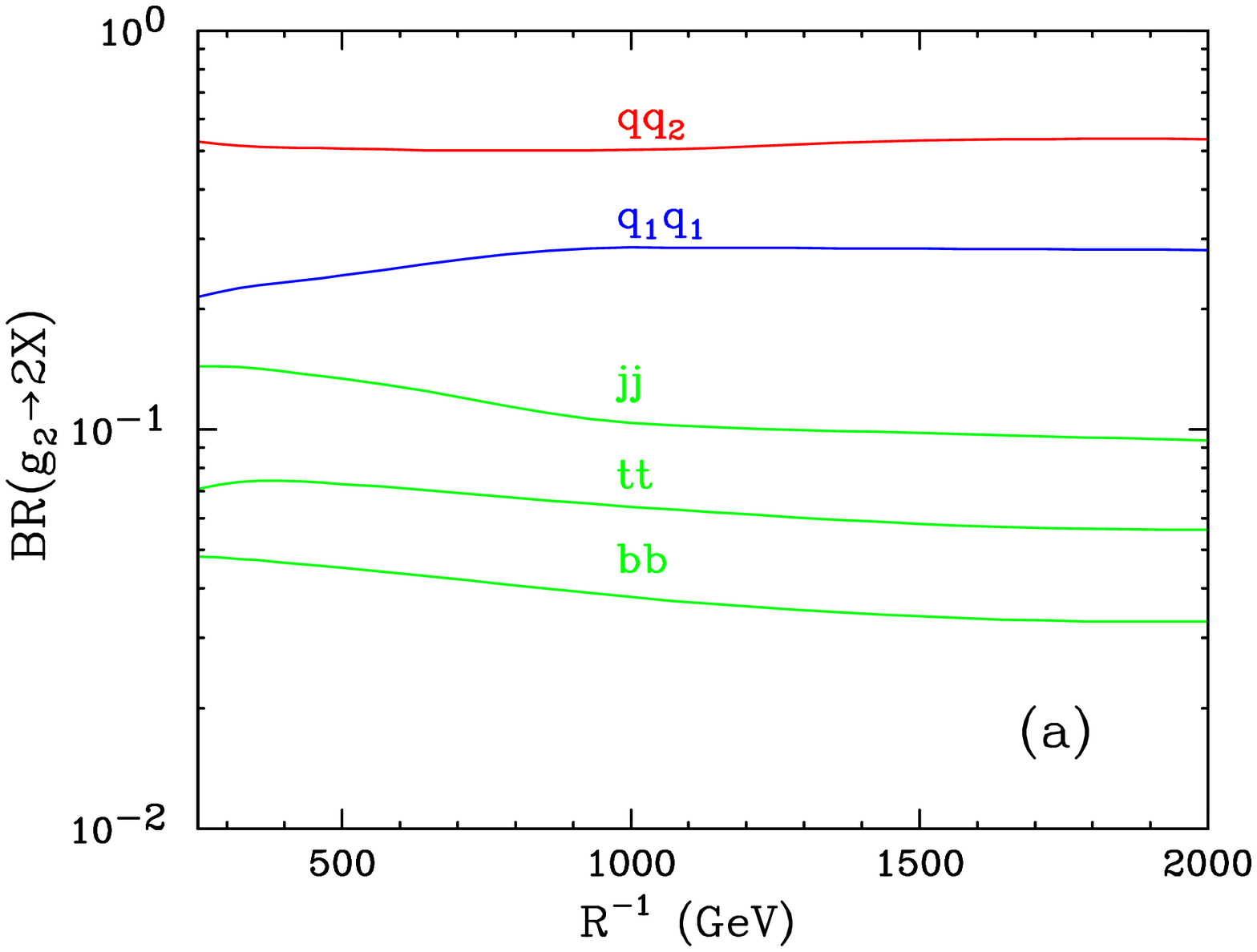}
\includegraphics[width=8cm]{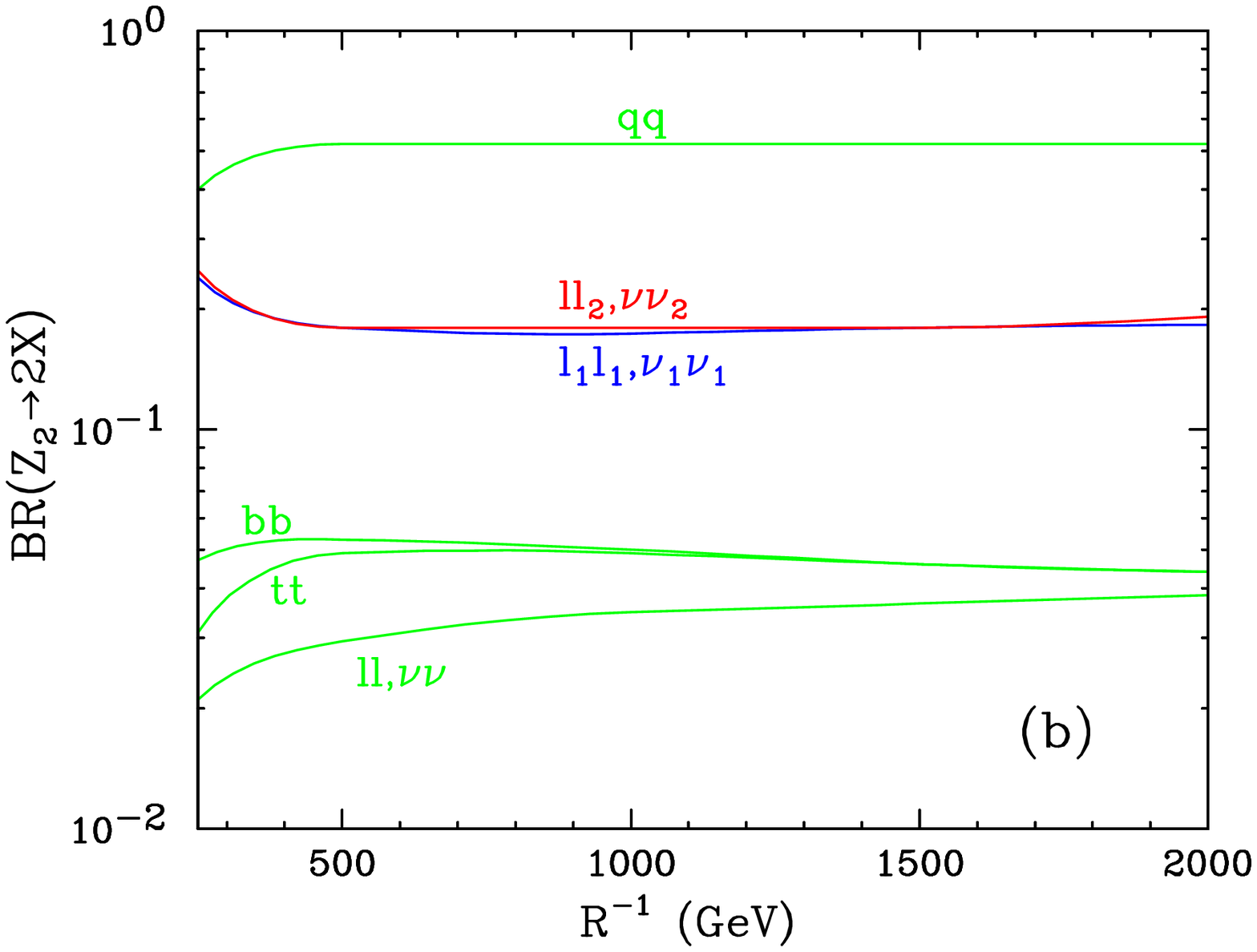}
\includegraphics[width=8cm]{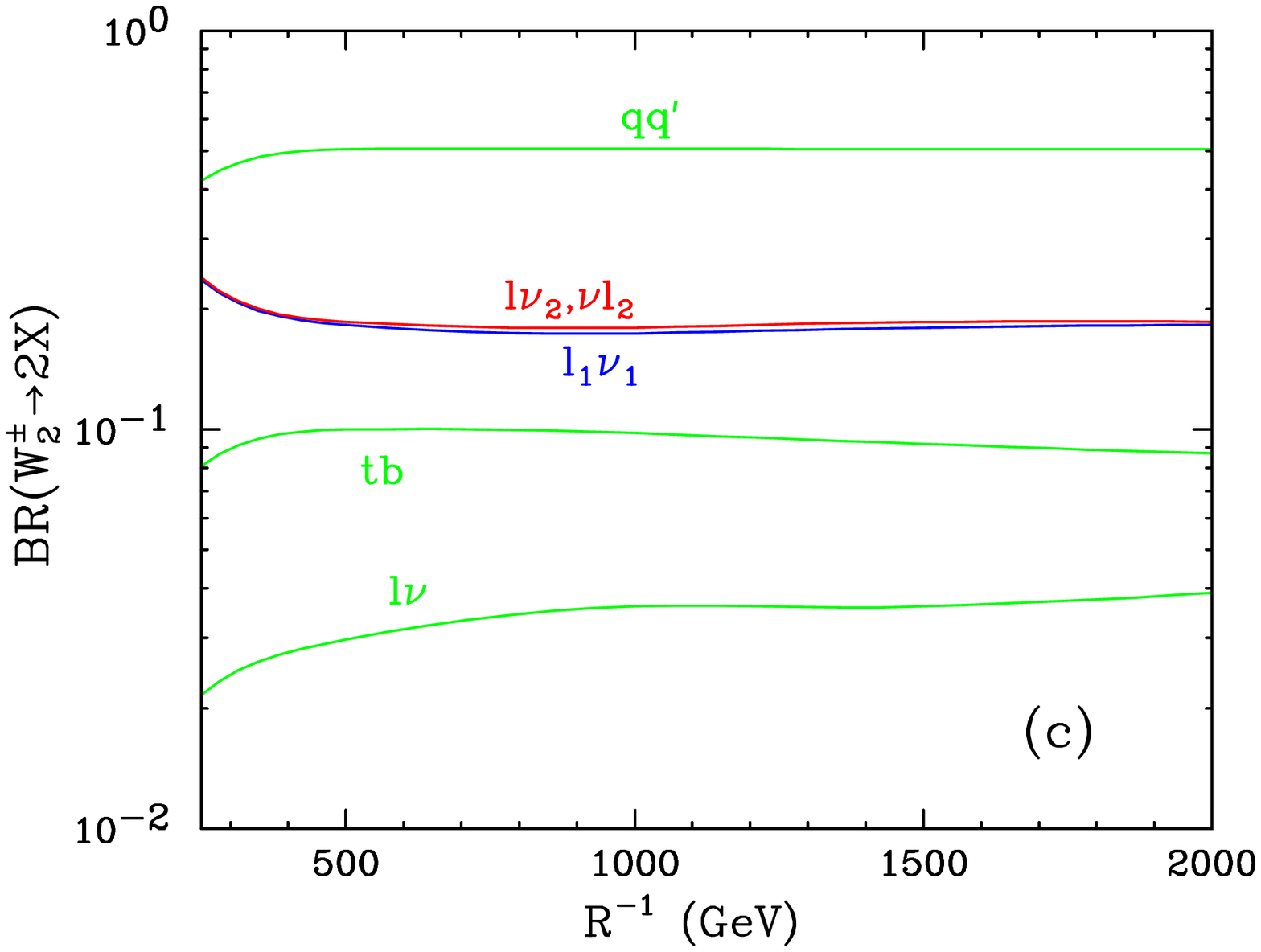}
\includegraphics[width=8cm]{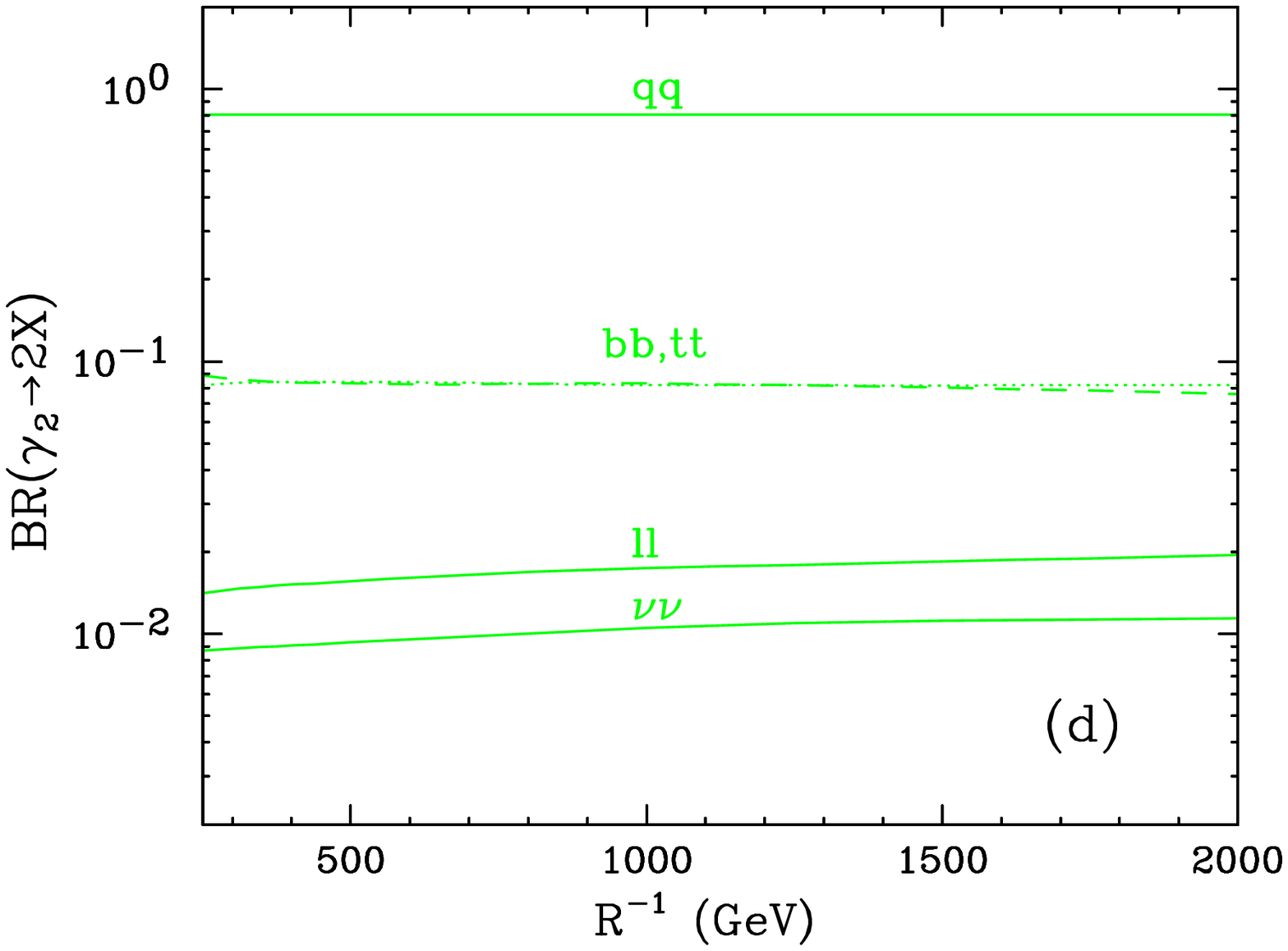}
\caption{Branching fractions of the $n=2$ KK gauge bosons
versus $R^{-1}$: (a) $g_2$, (b) $Z_2$, (c) $W^\pm_2$, and (d) $\gamma_2$.
\label{fig:br_V2}}
\end{figure}

In conclusion of this section, we discuss the experimental signatures 
of $n=2$ KK gauge bosons. To this end, we need to consider their possible 
decay modes. Having previously discussed the different partial widths, 
it is straightforward to compute the $V_2$ branching fractions.
Those are shown in Fig.~\ref{fig:br_V2}(a-d). Again we observe that
the branching fractions are very weakly sensitive to $R^{-1}$,
just as the case of Figs.~\ref{fig:br_q2} and \ref{fig:br_l2}.
This can be understood as follows. The partial widths (\ref{G220})
and (\ref{G211}) for the KK number conserving decays are proportional 
to the available phase space, while the partial width (\ref{G200})
for the KK number violating decay is proportional to the mass corrections
(see eq.~(\ref{gQ0Q0A2})). Both the phase space and the mass corrections 
are proportional to $R^{-1}$, which then cancels out in the branching fraction.

Similarly to the case
of $n=2$ KK quarks discussed in Sec.~\ref{sec:f2}, KK number conserving 
decays are not very distinctive, since they simply contribute to the 
inclusive $n=1$ sample which is dominated by direct $n=1$ production. 
The decays of $n=1$ particles will then give relatively soft objects, 
and most of the energy will be lost in the LKP mass. In short,
$n=2$ signatures based on purely KK number conserving decays are
not very promising experimentally --- one has to pay a big price in 
the cross-section in order to produce the heavy $n=2$ particles, but
does not get the benefit of the large mass, since most of the energy is
carried away by the invisible LKP. We therefore concentrate on the
KK number violating channels, in which the $V_2$ decays are fully visible.

Fig.~\ref{fig:br_V2}a shows the branching fractions of the KK gluon $g_2$.
Since it is the heaviest particle at level 2, all of its decay modes are 
open, and have comparable branching fractions. The KK number conserving
decays dominate, since the KK number violating coupling is slightly
suppressed due to the cancellation in (\ref{gQ0Q0A2}). In principle,
$g_2$ can be looked for as a resonance in the dijet~\cite{Dicus:2000hm} 
or $t\bar{t}$ invariant mass spectrum, but one would expect large backgrounds from 
QCD and Drell-Yan. Notice that there is no indirect production of $g_2$,
and its single production cross-section is not that much different
from the cross-sections for $\gamma_2$, $Z_2$ and $W^\pm_2$ 
(see Fig.~\ref{fig:sigma_V2}). Therefore, the inclusive $g_2$ production
is comparable to the inclusive $\gamma_2$ and $Z_2$ production, and then
we anticipate that the searches for the $n=2$ electroweak gauge bosons 
in leptonic channels will be more promising.

Figs.~\ref{fig:br_V2}b and \ref{fig:br_V2}c give the branching fractions of
$Z_2$ and $W^\pm_2$, correspondingly. We see that the decays to KK quarks 
have been closed due to the large QCD radiative corrections to the 
KK quark masses. Among the possible KK number conserving decays of $Z_2$ 
and $W^\pm_2$, only the leptonic modes survive, and they will be contributing 
to the leptonic discovery signals of UED~\cite{Cheng:2002ab}. 
Recall that the KK number conserving decays are phase space suppressed, 
while the KK number violating decays are loop suppressed, and 
proportional to the mass corrections as in (\ref{gQ0Q0A2}). 
The precise calculation shows that the dominant decay modes 
are $Z_2\to q\bar{q}$ and $W^\pm_2\to q\bar{q}'$. 
This can be understood in terms of the large $\bar\delta m_{q_2}$ 
correction appearing in (\ref{gQ0Q0A2}). 
The resulting branching ratios are more than $50\%$
and in principle allow for a $Z_2/W^\pm_2$ search in the dijet channel,
just like the case of $g_2$. However, we shall concentrate on the 
leptonic decay modes, which have much smaller branching fractions, 
but are much cleaner experimentally.

Finally, Fig.~\ref{fig:br_V2}d shows the branching fractions of
$\gamma_2$. This time all KK number conserving decays are closed, 
and $\gamma_2$ is forced to decay through the 
KK number violating interaction (\ref{Q0Q0A2}).
Again, the jetty modes dominate, and the leptonic modes 
(summed over lepton flavors) have rather small branching fractions, 
on the order of $2\%$, which could be a potential problem for the 
search. In the following section we shall concentrate
on the $Z_2\to \ell^+\ell^-$ and $\gamma_2\to \ell^+\ell^-$
signatures and analyze their discovery prospects 
in a $Z'$-like search~\cite{Nath:1999mw,Rizzo:2003ug}.

\subsection{Analysis of the LHC reach for $Z_2$ and $\gamma_2$}
\label{sec:analysis}

We are now in a position to discuss the discovery reach of
the $n=2$ KK gauge bosons at the LHC and the Tevatron.
We will consider the inclusive production of $Z_2$ and $\gamma_2$
and look for a dilepton resonance in both the $e^+e^-$ and $\mu^+\mu^-$ 
channels. An important parameter of the search is the
width of the reconstructed resonance, which in turn determines 
the size of the invariant mass window selected by the cuts.
Since the intrinsic width of the $Z_2$ and $\gamma_2$ resonances
is so small (see Fig.~\ref{fig:Gammas}b), the mass window is 
entirely determined by the mass resolution in the dimuon and 
dielectron channels. For electrons, the resolution in CMS 
is approximately constant, on the order of $\Delta m_{ee}/m_{ee}\approx 1\%$ 
in the region of interest~\cite{Darin}. On the other hand, the dimuon mass
resolution is energy dependent, and in preliminary studies based 
on a full simulation of the CMS detector has been parametrized 
as~\cite{muonres}
$$\frac{\Delta m_{\mu\mu}}{m_{\mu\mu}}=
0.0215+0.0128\left(\frac{m_{\mu\mu}}{1\ {\rm TeV}}\right)\ .$$
Therefore in our analysis we impose the following cuts 
\begin{enumerate}
\item Lower cuts on the lepton transverse momenta $p_T(\ell)>20$ GeV.
\item Central rapidity cut on the leptons $|\eta(\ell)|<2.4$.
\item Dilepton invariant mass cut for electrons
$m_{V_2}-2\Delta m_{ee}<m_{ee}<m_{V_2}+2\Delta m_{ee}$
and muons 
$m_{V_2}-2\Delta m_{\mu\mu}<m_{\mu\mu}<m_{V_2}+2\Delta m_{\mu\mu}$.
\end{enumerate}
With these cuts the signal efficiency
varies from $65\%$ at $R^{-1}=250$ GeV to $91\%$ at $R^{-1}=1$ TeV.
The main SM background to our signal is Drell-Yan,
which we have calculated with the {\tt PYTHIA} 
event generator~\cite{Sjostrand:2003wg}.

\begin{figure}[t]
\includegraphics[width=8cm]{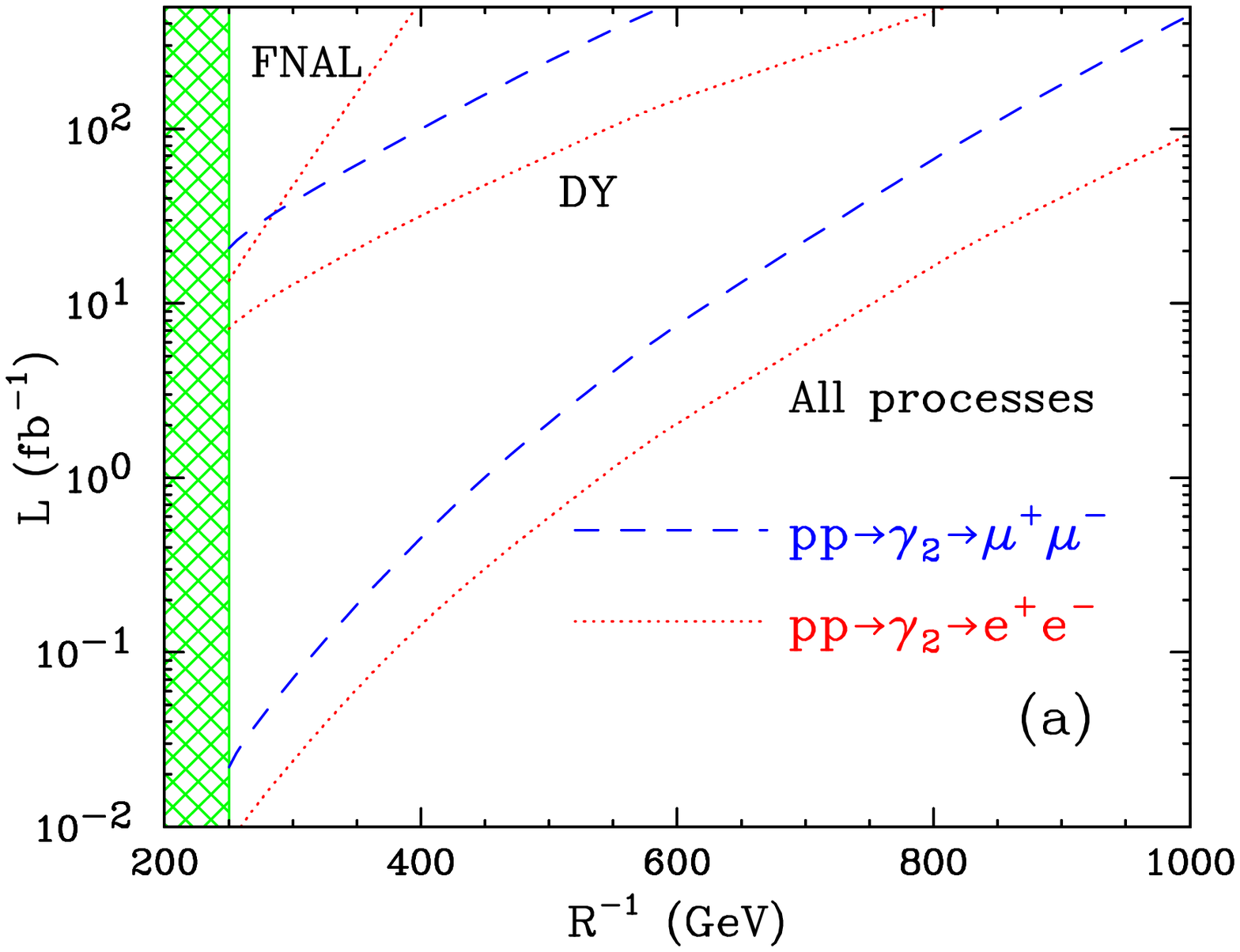}
\includegraphics[width=8cm]{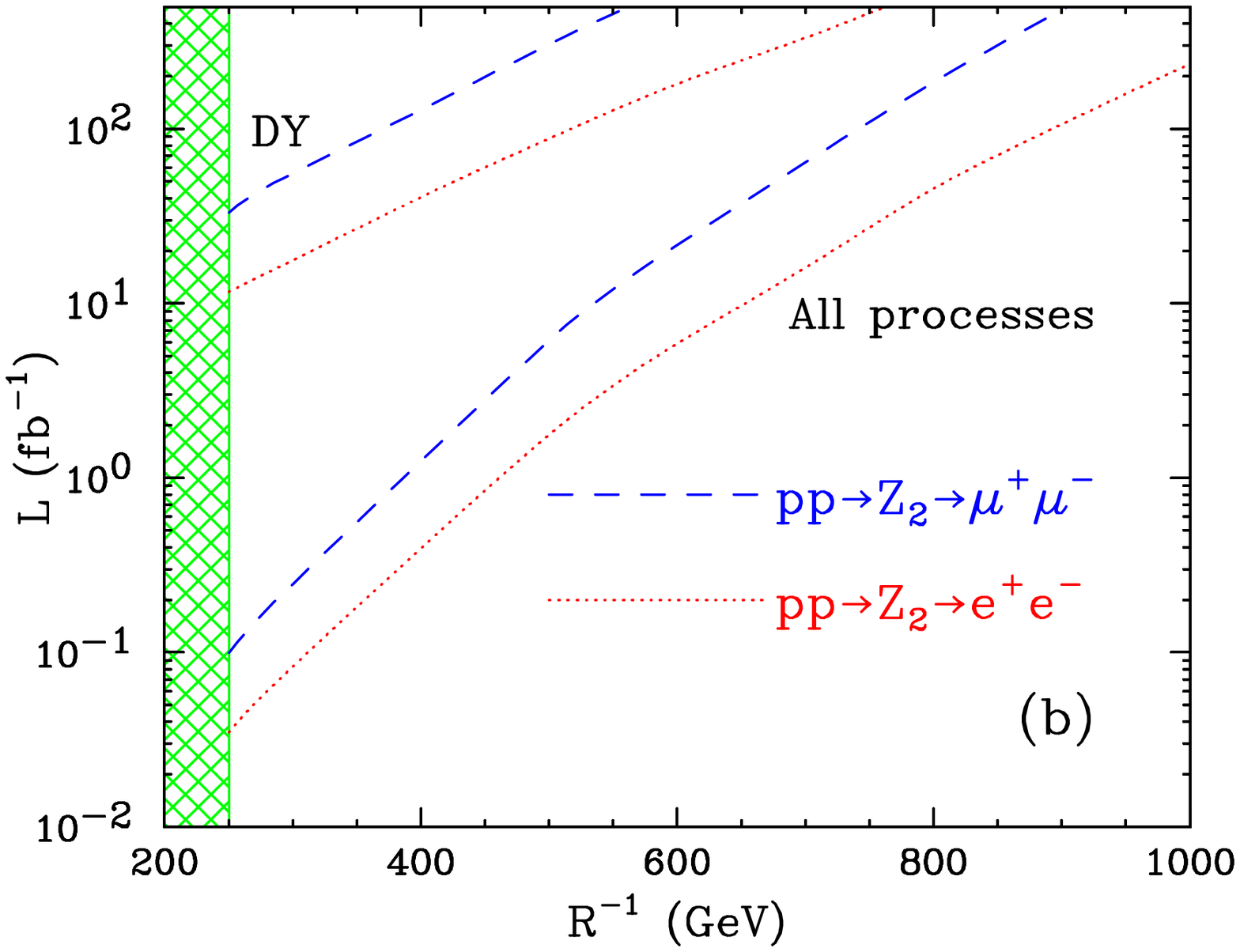}
\caption{$5\sigma$ discovery reach for (a) $\gamma_2$ and (b) $Z_2$.
We plot the total integrated luminosity ${\rm L}$ (in ${\rm fb}^{-1}$)
required for a $5\sigma$ excess of signal over background in the
dielectron (red, dotted) or dimuon (blue, dashed) channel,
as a function of $R^{-1}$. In each plot, the upper set of lines 
labelled ``DY'' makes use of the single $V_2$ production 
of Fig.~\ref{fig:sigma_V2} only,
while the lower set of lines (labelled ``All processes'')
includes indirect $\gamma_2$ and $Z_2$ production from 
$n=2$ KK quark decays. The red dotted line marked ``FNAL'' 
in the upper left corner of (a) reflects the expectations 
for a $\gamma_2\to e^+e^-$ discovery at the Tevatron in Run II.
The shaded area below $R^{-1}=250$ GeV indicates the
region disfavored by precision electroweak data~\cite{Appelquist:2002wb}.
\label{fig:reach}}
\end{figure}

With the cuts listed above, we compute the discovery reach 
of the LHC and the Tevatron for the $\gamma_2$ and $Z_2$ resonances.
Our results are shown in Fig.~\ref{fig:reach}.
We plot the total integrated luminosity ${\rm L}$ (in ${\rm fb}^{-1}$)
required for a $5\sigma$ excess of signal over background in the
dielectron (red, dotted) or dimuon (blue, dashed) channel,
as a function of $R^{-1}$. In each panel in Fig.~\ref{fig:reach}, 
the upper set of lines labelled ``DY'' only utilizes the 
single $V_2$ production cross-sections from Fig.~\ref{fig:sigma_V2}.
The lower set of lines (labelled ``All processes'')
includes in addition indirect $\gamma_2$ and $Z_2$ production from 
the decays of $n=2$ KK quarks to $\gamma_2$ and $Z_2$ (we ignore
secondary $\gamma_2$ production from $Q_2\to Z_2\to \ell_2 \to \gamma_2$). 
The shaded area below $R^{-1}=250$ GeV indicates the
region disfavored by precision electroweak data~\cite{Appelquist:2002wb}.
Using the same cuts also for the case of the Tevatron, we find 
the Tevatron reach in $\gamma_2\to e^+e^-$ shown in Fig.~\ref{fig:reach}a
and labelled ``FNAL''. For the Tevatron we use electron energy resolution 
$\Delta E/E=0.01\oplus0.16/\sqrt{E}$ \cite{Blair:1996kx}. 
The Tevatron reach in dimuons is worse 
due to the poorer resolution, while the reach for $Z_2$ is also worse
since $m_{Z_2}>m_{\gamma_2}$ for a fixed $R^{-1}$.

Fig.~\ref{fig:reach} reveals that there are good prospects for 
discovering level 2 gauge boson resonances at the LHC.
Already within one year of running at low luminosity 
(${\rm L}=10\ {\rm fb}^{-1}$), the LHC will have sufficient 
statistics in order to probe the region up to $R^{-1}\sim 750$ GeV.
Notice that in the Minimal UED model, the ``good dark matter'' region, 
where the LKP relic density accounts for all of the dark matter 
component of the Universe, is at $R^{-1}\sim 500-600$ GeV
\cite{Servant:2002aq,Burnell:2005hm,Kong:2005hn}. 
This region is well within the discovery 
reach of the LHC for both $n=1$ KK modes \cite{Cheng:2002ab}
and $n=2$ KK gauge bosons (Fig.~\ref{fig:reach}).
If the LKP accounts for only {\em a fraction of} the 
dark matter, the preferred range of $R^{-1}$ is even lower
and the discovery at the LHC is easier.

From Fig.~\ref{fig:reach} we also see that the ultimate reach of the 
LHC for both $\gamma_2$ and $Z_2$, 
after several years of running at high luminosity 
(${\rm L}\sim300\ {\rm fb}^{-1}$),
extends up to just beyond $R^{-1}=1$ TeV. One should keep in mind that
the actual KK masses are at least twice as large:
$m_{V_2}\sim m_2=2/R$, so that the KK resonances
can be discovered for masses up to $2$ TeV.

\begin{figure}[t]
\includegraphics[width=8cm]{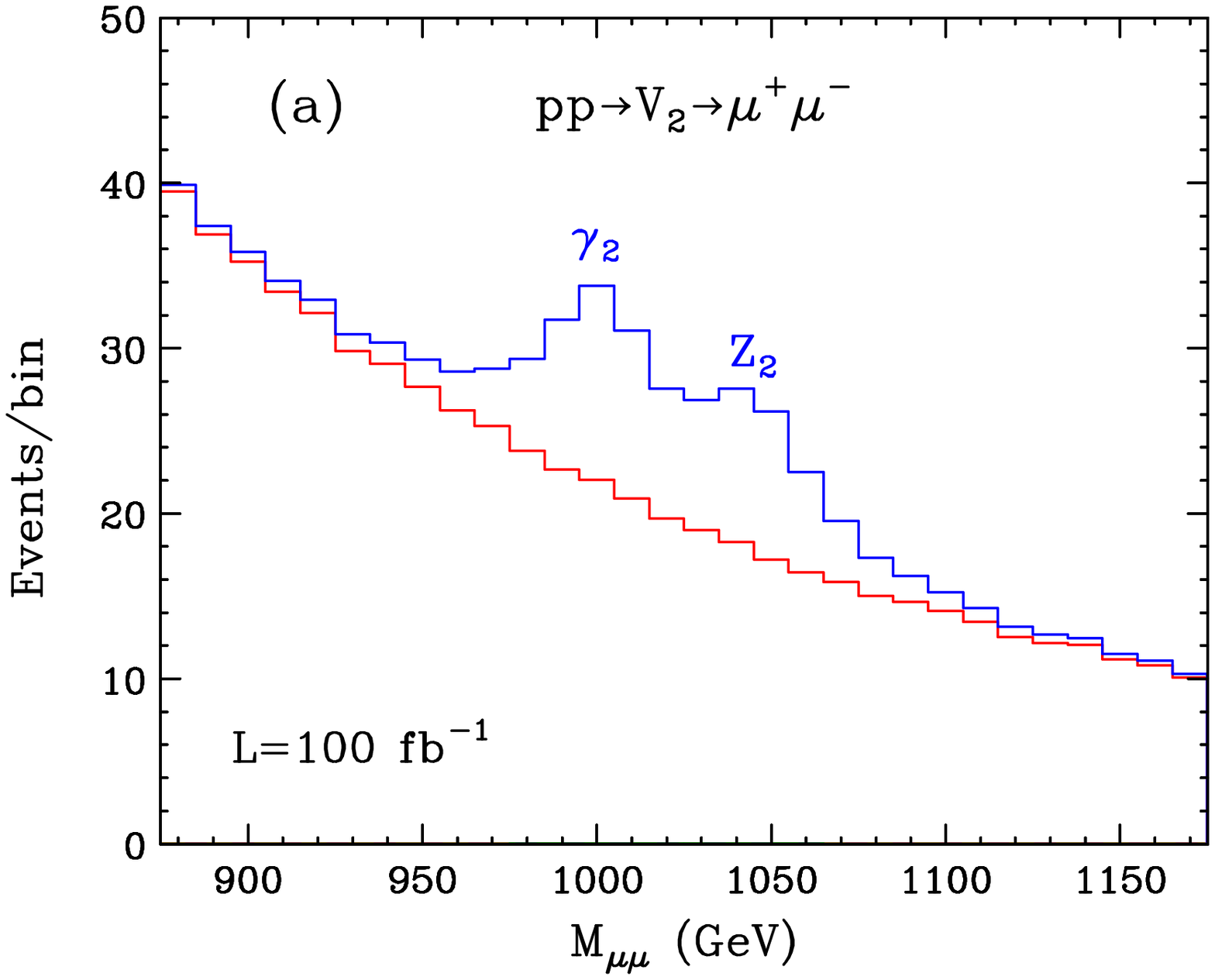}
\includegraphics[width=8cm]{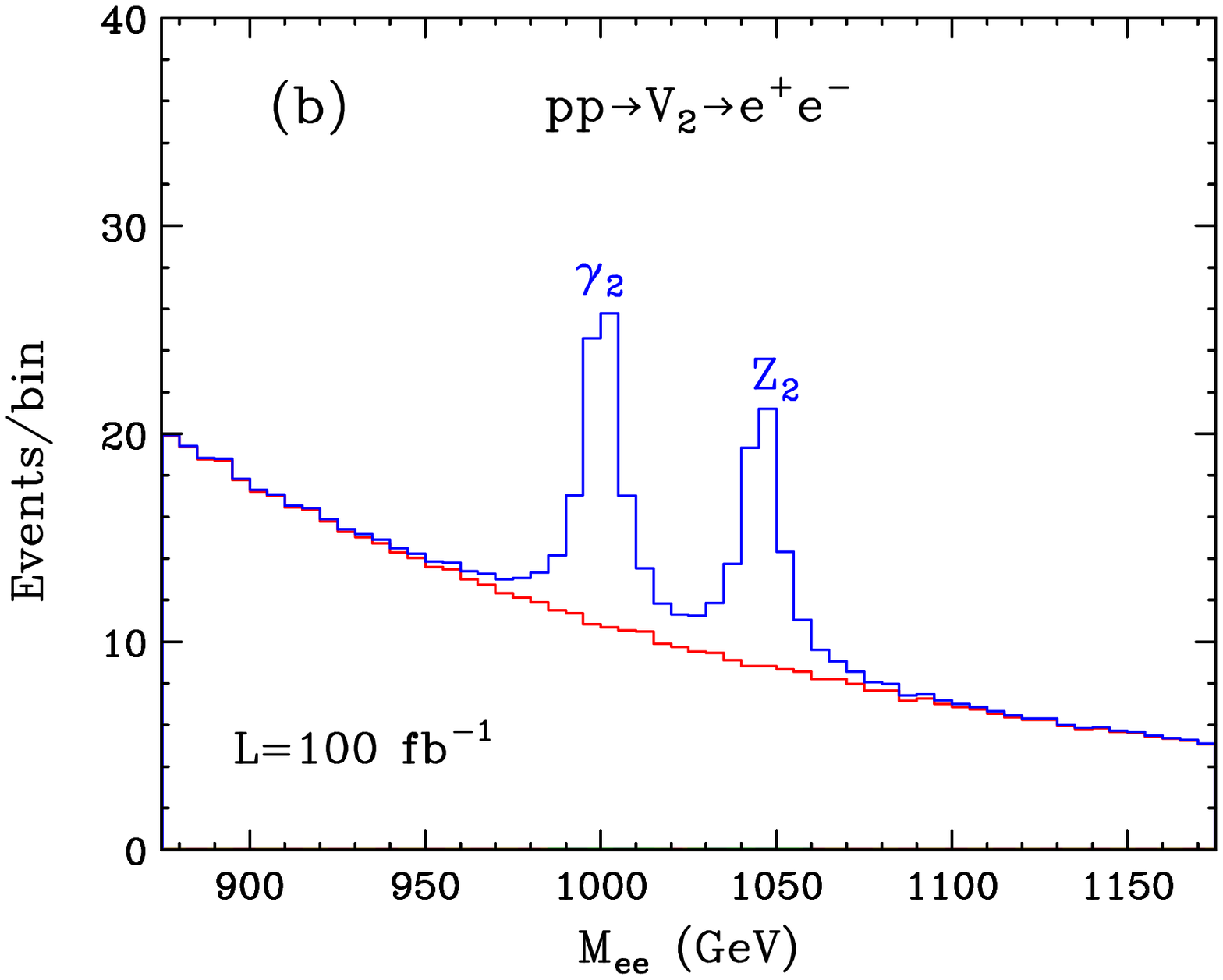}
\caption{The $\gamma_2-Z_2$ diresonance structure in 
UED with $R^{-1}=500$ GeV, for 
(a) the dimuon and (b) the dielectron channel at the LHC with 
${\rm L}=100\ {\rm fb}^{-1}$. The SM background is shown with the 
(red) continuous underlying histogram.
\label{fig:mll}}
\end{figure}

While the $n=2$ KK gauge bosons are a salient feature of the UED 
scenario, any such resonance by itself is not a sufficient discriminator,
since it resembles an ordinary $Z'$ gauge boson. If UED is discovered,
one could then still make the argument that it is in fact some sort of
non-minimal supersymmetric model with an additional gauge structure
containing neutral gauge bosons. An important corroborating evidence 
in favor of UED
would be the simultaneous discovery of several, rather degenerate,
KK gauge boson resonances. While SUSY also can accommodate multiple
$Z'$ gauge bosons, there would be no good motivation behind their
mass degeneracy. A crucial question therefore arises: can we separately 
discover the $n=2$ KK gauge bosons as individual resonances?
For this purpose, one would need to see a double peak structure
in the invariant mass distributions. Clearly, this is rather 
challenging in the dijet channel, due to the relatively poor 
jet energy resolution. We shall therefore consider only the dilepton 
channels, and investigate how well we can separate $\gamma_2$ from $Z_2$.

Our results are shown in Fig.~\ref{fig:mll}, where we show the invariant 
mass distribution in UED with $R^{-1}=500$ GeV, for 
(a) the dimuon and (b) the dielectron channel 
at the LHC with ${\rm L}=100\ {\rm fb}^{-1}$. We see that 
the diresonance structure is easier to detect in the dielectron channel, 
due to the better mass resolution. In dimuons, with ${\rm L}=100\ {\rm fb}^{-1}$
the structure is also beginning to emerge. We should note that 
initially the two resonances will not be separately distinguishable, 
and each will in principle contribute to the discovery of a bump,
although with a larger mass window. In our reach plots in Fig.~\ref{fig:reach}
we have conservatively chosen not to combine the two signals 
from $Z_2$ and $\gamma_2$, but show the reach for each one separately.

\section{\label{sec:spin}Spin Discriminations in SUSY and UED}

As discussed in Section~\ref{sec:intro}, the
second fundamental distinction between UED and supersymmetry is 
reflected in the properties of the individual particles: the
KK partners have identical spin quantum numbers as their SM counterparts, 
while the spins of the superpartners differ by $1/2$ unit. 
However, spin determinations appear to be difficult at the LHC
(or at hadron colliders in general), where the center of mass energy
in each event is unknown. In addition, the momenta of the two dark
matter candidates in the event are also unknown. Recently it has been
suggested that a charge asymmetry in the lepton-jet invariant mass
distributions from a particular cascade, can be used to discriminate
SUSY from the case of pure phase space decays~\cite{Barr:2004ze}
and is an indirect indication of the superparticle spins.
It is therefore natural to ask whether this method can be
extended to the case of SUSY versus UED discrimination.

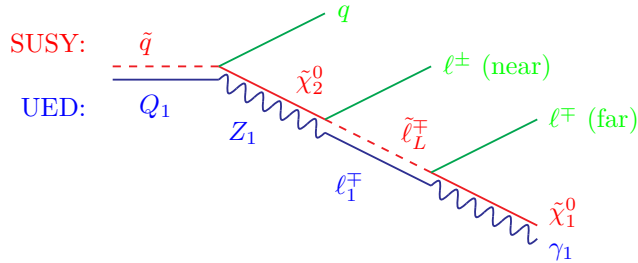
\begin{figure*}[t]
\begin{center}
\unitlength=1.0 pt
\SetScale{1.0}
\SetWidth{0.7}      
\footnotesize    
%
\begin{picture}(200,100)(0,0)
\SetColor{Red}
\Text(  0.0,79.0)[r]{\red{SUSY:}}
\Text( 20.0,79.0)[l]{\red{$\tilde q$}}
\Text( 90.0,65.0)[r]{\red{$\tilde \chi^0_2$}}
\Text(130.0,45.0)[r]{\red{$\tilde \ell^\mp_L$}}
\Text(175.0,15.0)[l]{\red{$\tilde \chi^0_1$}}
\DashLine(10.0,70.0)(50.0,70.0){3}
\Line(50.0,70.0)(90.0,50.0)
\DashLine(90.0,50.0)(130.0,30.0){3}
\Line(130.0,30.0)(170.0,10.0)
\SetColor{Blue}
\Text(  0.0,55.0)[r]{\blue{UED:}}
\Text( 20.0,55.0)[l]{\blue{$Q_1$}}
\Text( 65.0,45.0)[r]{\blue{$Z_1$}}
\Text(105.0,25.0)[r]{\blue{$\ell^\mp_1$}}
\Text(175.0, 0.0)[l]{\blue{$\gamma_1$}}
\Line(10.0,65.0)(50.0,65.0)
\Photon(50.0,65.0)(90.0,45.0){3}{6}
\Line(90.0,45.0)(130.0,25.0)
\Photon(130.0,25.0)(170.0,5.0){3}{6}
\SetColor{Green}
\Text( 95.0,90.0)[l]{\green{$q$}}
\Text(135.0,70.0)[l]{\green{$\ell^\pm$ (near)}}
\Text(175.0,50.0)[l]{\green{$\ell^\mp$ (far)}}
\Line( 50.0,70.0)( 90.0,90.0)
\Line( 90.0,50.0)(130.0,70.0)
\Line(130.0,30.0)(170.0,50.0)
\end{picture}
\end{center}
\caption{Twin diagrams in SUSY and UED. The upper (red) line corresponds 
to the cascade decay $\tilde q \to q\tilde\chi^0_2 \to q\ell^\pm\tilde\ell^\mp_L
\to q\ell^+\ell^-\tilde\chi^0_1$ in SUSY. The lower (blue) line corresponds 
to the cascade decay $Q_1 \to q Z_1\to q\ell^\pm\ell^\mp_1
\to q\ell^+\ell^-\gamma_1$ in UED. In either case the observable 
final state is the same: $q\ell^+\ell^-\met$.}
\label{fig:diagrams}
\end{figure*}

To answer this question, we first choose a study point in UED with
$R^{-1}=500$ GeV. Then we adjust the relevant MSSM parameters until
we get a matching spectrum. Following~\cite{Barr:2004ze}, we concentrate on
the cascade decay $\tilde q \to q\tilde\chi^0_2 \to q\ell^\pm\tilde\ell^\mp_L
\to q\ell^+\ell^-\tilde\chi^0_1$ in SUSY and the analogous decay chain
$Q_1 \to q Z_1\to q\ell^\pm\ell^\mp_1\to q\ell^+\ell^-\gamma_1$ in 
UED~\cite{KMAPS, KongPheno}.
Both of these processes are illustrated in Fig.~\ref{fig:diagrams}.

\begin{figure}[ht]
\includegraphics[width=8cm]{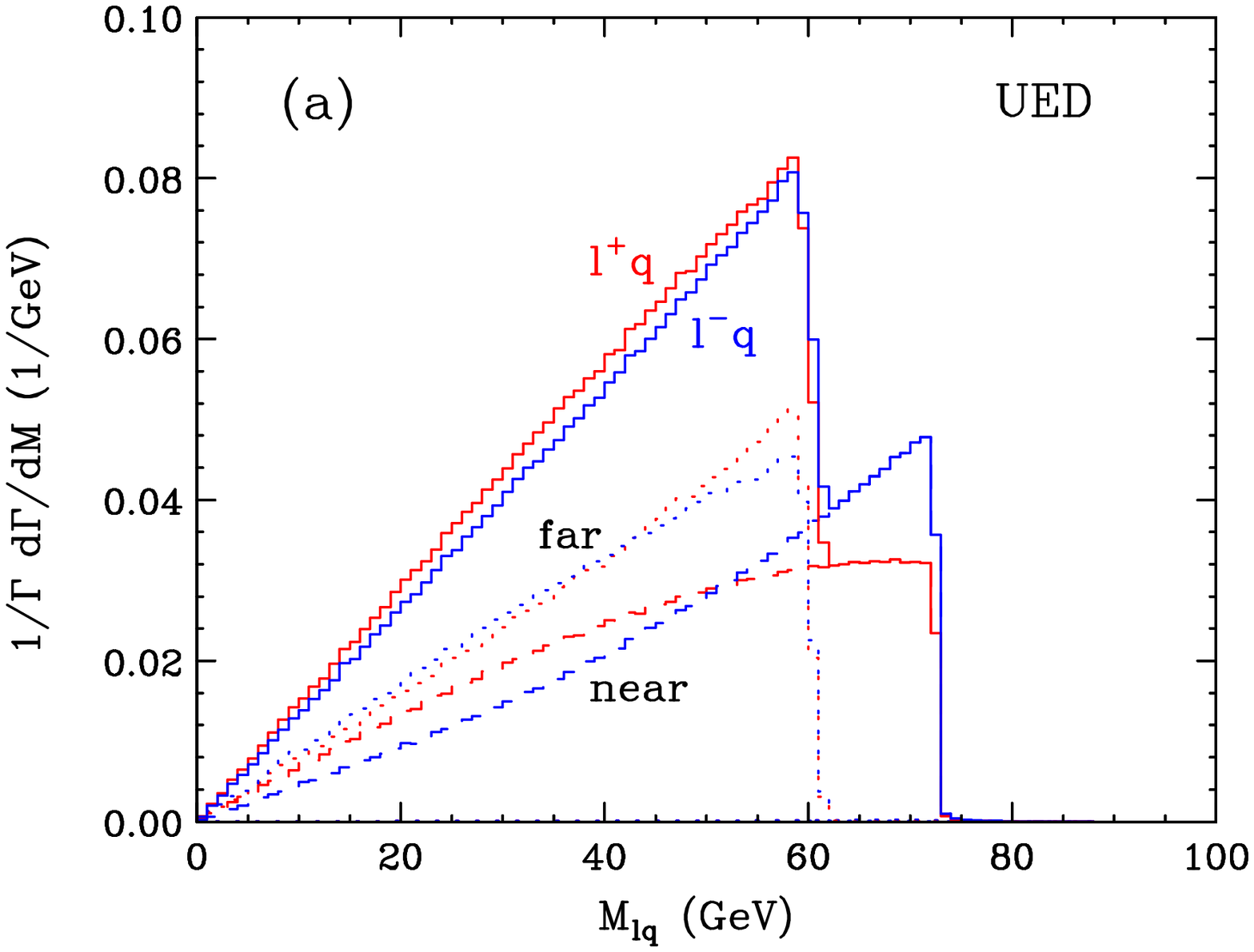}
\includegraphics[width=8cm]{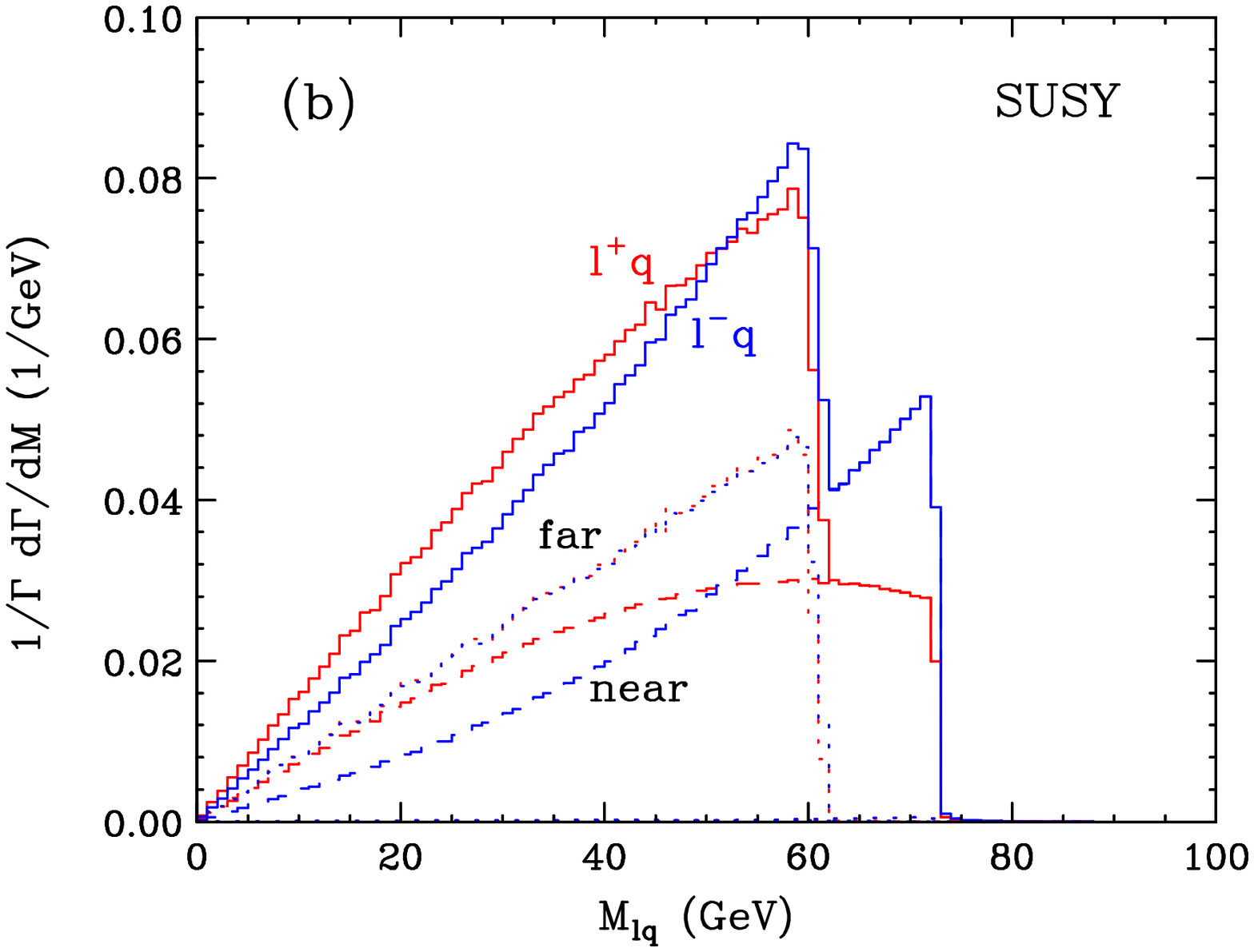}
\caption{Lepton-quark invariant mass distributions in 
(a) UED with $R^{-1}=500$ GeV and (b) supersymmetry with
a matching sparticle spectrum. We show separately the 
distributions with the near and far lepton, and their sum.
The positive (negative) charge leptons are shown in red (blue).
\label{fig:mql}}
\end{figure}
\begin{figure}[t]
\includegraphics[width=10cm]{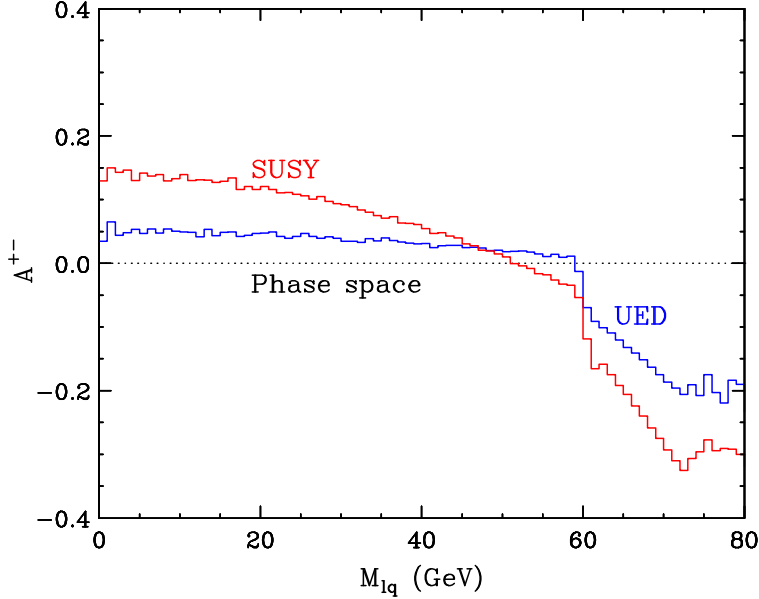}
\caption{Comparison of the charge asymmetry $A^{+-}$ defined in eq.~(\ref{asymmetry})
as computed in the case of UED with $R^{-1}=500$ GeV and the case of supersymmetry with
a matching sparticle spectrum.
\label{fig:asymmetry}}
\end{figure}

Next, one forms the lepton-quark invariant mass distributions 
$M_{\ell q}$ (see Fig.~\ref{fig:mql}). The spin of the intermediate particle 
($Z_1$ in UED or $\tilde\chi^0_2$ in SUSY) governs the shape
of the distributions for the near lepton.
However, in practice we cannot distinguish the near and far 
lepton, and one has to include the invariant mass combinations 
with both leptons. This tends to wash out the spin correlations, 
but a residual effect remains, which is due to the different 
number of quarks and antiquarks in the proton,
which in turn leads to a difference in the production cross-sections
for squarks and anti-squarks~\cite{Barr:2004ze}.
The spin correlations are encoded in the charge asymmetry~\cite{Barr:2004ze}
\begin{equation}
A^{+-} \equiv
\left(\frac{dN(q\ell^+)}{dM_{ql}}-\frac{dN(q\ell^-)}{dM_{ql}}\right)\Biggr/
\left(\frac{dN(q\ell^+)}{dM_{ql}}+\frac{dN(q\ell^-)}{dM_{ql}}\right)\ ,
\label{asymmetry}
\end{equation}
where $q$ stands for both a quark and an antiquark, and $N(q\ell^+)$
($N(q\ell^-)$) is the number of entries with positively (negatively) 
charged lepton. Our comparison 
between $A^{+-}$ in the case of UED and SUSY~\cite{KMAPS, KongPheno} 
is shown in Fig.~\ref{fig:asymmetry}. 
We see that although there is some minor difference in the shape of the 
asymmetry curves, overall the two cases appear to be very difficult to
discriminate unambiguously, especially since the regions near the two ends of the
plot, where the deviation is the largest, also happen to suffer from
poorest statistics. Notice that we have not included detector effects
or backgrounds. Finally, and perhaps most importantly, this analysis 
ignores the combinatorial background from the other jets in the event,
which could be misinterpreted as the starting point of the cascade 
depicted in Fig.~\ref{fig:diagrams}. Overall, Fig.~\ref{fig:asymmetry}
shows that although the asymmetry (\ref{asymmetry}) does encode some 
spin correlations, distinguishing between the specific cases of UED and SUSY 
appears challenging. These results have been recently confirmed 
in~\cite{Smillie:2005ar}, where in addition the authors considered 
a study point with larger mass splittings, as expected in typical 
SUSY models. Under those circumstances the asymmetry distributions 
appear to be more distinct than the case shown in Fig.~\ref{fig:asymmetry},
which is a source of optimism. It remains to be seen whether this 
conclusion persists in a general setting, and once the combinatorial 
backgrounds are included~\cite{DKM}.

\section{\label{sec:conclusions}Conclusions}

In this paper we have discussed the differences and
similarities in the hadron collider phenomenology of 
models with Universal Extra Dimensions and supersymmetry.
We identified the higher level KK modes of UED and the
spin quantum numbers of the new particles as the only two
reliable discriminators between the two scenarios. We then 
proceeded to study the discovery reach for level 2 KK modes
in UED at hadron colliders. We showed that the $n=2$ KK gauge 
bosons offer the best prospects for detection, in particular the 
$\gamma_2$ and $Z_2$ resonances can be {\em separately} 
discovered at the LHC. Is this a proof of UED? Not quite --
these resonances could still be interpreted as $Z'$ gauge bosons, 
but their close degeneracy is a smoking gun for UED.
Furthermore, although we did not show any results to 
this effect in this paper, it is clear that the $W_2^\pm$ 
KK mode can also be looked for and discovered in its 
decay to SM leptons. One can then measure $m_{W_2}$ and 
show that it is very close to $m_{Z_2}$ and $m_{\gamma_2}$,
which would further strengthen the case for UED.
Unfortunately, the spin discrimination is not so straightforward,
and requires further studies. The asymmetry method of Barr 
seems to fail as a universal discriminator between SUSY and UED, 
although it rules out the absence of any spin correlations.

While in this paper we only concentrated on the Minimal UED model,
it should be kept in mind that there are many interesting possibilities 
for extending the analysis to a more general setup.
For example, non-vanishing boundary terms at the scale $\Lambda$
can distort the Minimal UED spectrum beyond recognition.
A priori, in such a relaxed framework the UED-SUSY
confusion can be ``complete'' in the context of a hadron collider
and a preliminary study is under way to address this issue~\cite{DKT}.
The UED collider phenomenology is also very different 
in the case of a ``fat'' brane~\cite{Macesanu:2002ew,Macesanu:2004nb},
charged LKPs~\cite{Byrne:2003sa} or KK graviton superwimps
\cite{Feng:2003xh,Feng:2005gj}. Notice that Little Higgs models 
with $T$-parity \cite{Cheng:2003ju,Cheng:2004yc,Hubisz:2004ft}
are very similar to UED, and can also be confused with supersymmetry.

\begin{acknowledgments}
We thank H.-C.~Cheng and B.~Dobrescu for stimulating discussions.
AD is supported by the US Department of Energy and the 
Michigan Center for Theoretical Physics.
The work of KK and KM is supported in part by 
a US Department of Energy Outstanding Junior Investigator 
award under grant DE-FG02-97ER41209.
\end{acknowledgments}



\end{document}